\documentclass[12pt,preprint]{aastex}

 \usepackage{epstopdf}
 \usepackage{multirow}
\usepackage{float}

 \slugcomment{Accepted}

\newcommand\pv{\mbox{$p_{V}$}}
\newcommand\pIR{\mbox{$p_{IR}$}}

\newcommand\Vaisala{V\"{a}is\"{a}l\"{a}}

\begin{document}

 \DeclareGraphicsExtensions{.pdf,.gif,.jpg}

 \title{Survey Simulations of a New Near-Earth Asteroid Detection System}

\author{A. Mainzer\altaffilmark{1}, T. Grav\altaffilmark{2},J. Bauer\altaffilmark{1,3}, T. Conrow\altaffilmark{3}, R. M. Cutri\altaffilmark{3},  J. Dailey\altaffilmark{3}, J. Fowler\altaffilmark{3}, J. Giorgini\altaffilmark{1}, T. Jarrett\altaffilmark{3,4},  J. Masiero\altaffilmark{1}, T. Spahr\altaffilmark{5}, T. Statler\altaffilmark{6,7}, E. L. Wright\altaffilmark{8} }

\altaffiltext{1}{Jet Propulsion Laboratory, California Institute of Technology, Pasadena, CA 91109 USA}
\altaffiltext{2}{Planetary Science Institute, Tucson, AZ USA}
\altaffiltext{3}{Infrared Processing and Analysis Center, California Institute of Technology, Pasadena, CA 91125, USA}
\altaffiltext{4}{Astronomy Department, University of Cape Town, Rondebosch 7701 South Africa}
\altaffiltext{5}{Minor Planet Center, Harvard-Smithsonian Center for Astrophysics, 60 Gardent St. MS-18, Cambridge, MA 02138 USA}
\altaffiltext{6}{Astrophysical Institute, Ohio University}
\altaffiltext{7}{Department of Astronomy, University of Maryland, College Park}
\altaffiltext{8}{Department of Astronomy and Astrophysics, University of California Los Angeles, Los Angeles, CA USA}


 \begin{abstract}
We have carried out simulations to predict the performance of a new space-based telescopic survey operating at thermal infrared wavelengths that seeks to discover and characterize a large fraction of the potentially hazardous near-Earth asteroid (NEA) population.  Two potential architectures for the survey were considered: one located at the Earth-Sun L1 Lagrange point, and one in a Venus-trailing orbit.  A sample cadence was formulated and tested, allowing for the self-follow-up necessary for objects discovered in the daytime sky on Earth.  Synthetic populations of NEAs with sizes as small as 140 m in effective spherical diameter were simulated using recent determinations of their physical and orbital properties.  Estimates of the instrumental sensitivity, integration times, and slew speeds were included for both architectures assuming the properties of newly developed large-format 10 $\mu$m HgCdTe detector arrays capable of operating at $\sim$35 K.  Our simulation included the creation of a preliminary version of a moving object processing pipeline suitable for operating on the trial cadence. We tested this pipeline on a simulated sky populated with astrophysical sources such as stars and galaxies extrapolated from \emph{Spitzer Space Telescope} and \emph{Wide-field Infrared Explorer} data, the catalog of known minor planets (including Main Belt asteroids, comets, Jovian Trojans, planets, etc.), and the synthetic NEA model.  Trial orbits were computed for simulated position-time pairs extracted from the synthetic surveys to verify that the tested cadence would result in orbits suitable for recovering objects at a later time.  Our results indicate that the Earth-Sun L1 and Venus-trailing surveys achieve similar levels of integral completeness for potentially hazardous asteroids larger than 140 m; placing the telescope in an interior orbit does not yield an improvement in discovery rates.  This work serves as a necessary first step for the detailed planning of a next-generation NEA survey.

 \end{abstract}

 \section{Introduction}
From the kilometer-diameter Meteor Crater to the enormous ring-shaped feature beneath the Gulf of Mexico resulting from the Cretaceous-Paleogene impactor\citep{Alvarez.1980a,Hildebrand.1991a}, numerous lines of evidence record the history of previous near-Earth object (NEO) impacts.  More recent events, such as comet Shoemaker-Levy 9's impact with Jupiter \citep{Levy.1995a}, have spurred international interest in observational efforts to determine the collisional hazard to Earth posed by NEOs.  To date, most efforts have focused on ground-based surveys carried out by 0.5 to 2 m class telescopes at visible wavelengths \citep[e.g. the Near Earth Asteroid Tracking Program, the Lincoln Near-Earth Asteroid Research Program, the Lowell Near-Earth Object Survey, Spacewatch, Pan-STARRS, and the Catalina Sky Survey;][]{Helin.1997a,Stokes.2000a,Koehn.2000a,Larson.2007a,McMillan.2007a}.  These successful programs have now discovered 90\% of all NEOs larger than 1 km in diameter \citep{Mainzer.2011b}.  A total of $\sim$12,000 NEOs have been found to date (\emph{http://neo.jpl.nasa.gov/stats.html}).  Dedicated systems at the International Astronomical Union's Minor Planet Center (MPC) archive and link new observations of NEOs in real time. Impact probabilities are updated in real time by the NASA Near Earth Object Program's SENTRY system and the European Union's NEODys system\footnote{\emph{http://neo.jpl.nasa.gov/risk/}; \emph{http://newton.dm.unipi.it/neodys/}} \citep{Milani.2005a}.

The $\sim$12,000 NEOs known so far are believed to represent only a small fraction of the total population.  Statistical approaches to quantifying the hazard from the undiscovered population indicate that while the risk of a very large impact is low over the next several thousand years, dangerous NEO impacts remain a stochastic process on human timescales.  In 2004, the U. S. Congress mandated that NASA should find more than 90\% of all NEOs larger than 140 m in diameter by the year 2020.  The process of surveying for NEOs has proven extremely beneficial to planetary science.  The survey data used to discover, catalog and characterize NEOs are responsible for finding the majority of comets, Main Belt asteroids, Jovian Trojans, and Centaurs known today, a tremendous boon to those interested in understanding the origins and evolution of our solar system.  

Additionally, the NEOs that are likely to be the most hazardous tend to make numerous close approaches to the Earth, making some of them potentially attractive destinations for future human and robotic missions.  NASA has been charged by the President of the United States with a new initiative to send human explorers to a near-Earth asteroid in the 2025 timeframe \citep[c.f.][]{Abell.2009a}.  Out of the NEOs known to date, only a handful meet requirements for size and accessibility in the timeframe of interest with sufficiently short trip durations (http://neo.jpl.nasa.gov/nhats/).  Little is known about these few objects in most cases other than their orbits, which are often highly uncertain, and their absolute visible magnitudes.  NEO surveys are needed to find and characterize suitable targets for exploration, to identify and characterize potentially hazardous objects, and to carry out larger-scale scientific studies of the small body population of our solar system.  All of these reasons motivate efforts to carry out a survey capable of not only discovering a large fraction of NEOs big enough to cause major damage but also to systematically characterize their physical and orbital properties.  

There is precedent for using space-based IR telescopes to discover and characterize asteroids in large numbers.  NEOs emit a significant fraction of their bolometric luminosity at thermal infrared (IR) wavelengths, and the stellar and galactic backgrounds against which they are detected are typically much dimmer than in visible light.  NEO thermal emission is only a weak function of an object's geometric albedo, unlike its visible light brightness, so a telescope surveying at IR wavelengths is essentially equally sensitive to low and high albedo objects \citep{Mainzer.2011b}.  Moreover, surveying at IR wavelengths offers the opportunity to constrain an object's diameter with much improved accuracy over visible wavelengths alone.    Independent IR surveys are therefore useful when extrapolating the observed sample to determine properties of the population such as size frequency distributions.  No conversion between $H$ and diameter is needed, since diameter is determined directly, eliminating the uncertainty associated with the large range of possible asteroid albedos.  If both visible and IR fluxes are obtained, it is possible to compute geometric albedo \citep{Harris.1998a, Mainzer.2011c, Mainzer.2011d}.  Finally, IR wavelengths can be used to set constraints on the properties of regolith such as thermal inertia and conductivity \citep[e.g.][]{Ali-Lagoa.2014a, Groussin.2011a, Delbo.2007a}.  

NEOWISE, the asteroid-hunting portion of NASA's \emph{Wide-field Infrared Survey Explorer} \citep[WISE;][]{Wright.2010a} mission, discovered and characterized $>$158,000 asteroids and comets throughout our solar system during its year-long prime mission \citep{Mainzer.2011a}; $>$34,000 were new discoveries.  Included in the sample were $\sim$700 NEOs, of which 135 were new discoveries \citep{Mainzer.2011b, Mainzer.2012b, Mainzer.2014a}.  NEOWISE discovered the first known Earth Trojan, 2010 TK$_{7}$ \citep{Connors.2011a} as well as a so-called ``horseshoe" NEO \citep[2010 SO$_{16}$;][]{Christou.2011a}.  By virtue of its wavelengths, all-sky coverage, and uniformity, the NEOWISE sample has been used to set constraints on the numbers, sizes, and orbits of asteroids, including NEOs and those that are classified as potentially hazardous \citep{Mainzer.2012a, Grav.2011b, Grav.2012a}.  NEOWISE data have been used to study the size and albedo distributions of asteroids throughout the Main Belt \citep[e.g.][]{Masiero.2011a, DeMeo.2014a}, and to identify new collisional family members using albedo as an additional constraint \citep{Masiero.2013a, Carruba.2013a, Walsh.2013a, Milani.2014a}.  After a 32 month hibernation following the completion of its prime mission in February 2011, the spacecraft was brought back into operation, renamed NEOWISE, and has resumed discovering and characterizing NEOs using its remaining 3.4 and 4.6 $\mu$m channels \citep{Mainzer.2014b}.

Other space missions have been used to study asteroids and comets, with smaller numbers of new discoveries.  The \emph{Infrared Astronomical Satellite} \citep[IRAS;][]{Neugebauer.1984a, Beichman.1988a} surveyed the sky and observed $\sim$2000 small bodies, mostly previously known Main Belt asteroids \citep{Tedesco.2002a}.  Additional observations came from the \emph{Mid-Course Space Experiment} \citep[MSX;][]{Mill.1994a, Price.2001a, Tedesco.2002b}.  The \emph{Spitzer Space Telecope} has observed hundreds of previously known NEOs \citep[e.g.][]{Trilling.2010a, Mueller.2011a, Thomas.2011a}. The AKARI mission observed $\sim$5000 previously known asteroids \citep{Usui.2011a, Usui.2013a, Hasegawa.2013a}. 

Using a space-based telescope operating at thermal IR wavelengths to discover and characterize large numbers of NEOs is not a new concept.  \citet{Tedesco.2000a}, \citet{Cellino.2000a}, \citet{Cellino.2003a}, and \citet{Cellino.2004a} considered space-based thermal IR telescopes observing from Earth orbit, the Earth-Sun L2 Lagrange point, and a Venus-like orbit using HgCdTe detectors cooled by mechanical cryocoolers.  However, the 2003 report of the NASA NEO Science Definition Team \citep{Stokes.2003a} opted not to consider a space-based IR system on the grounds that sufficiently large, high-operability long-wavelength focal plane arrays did not exist at the time.  At the time, the largest long-wavelength ($\sim10 \mu$m) HgCdTe detectors designed for low-background astronomical background applications were in a 512x512 format, and their dark current and operability were inadequate for natural background-limited performance.  Nevertheless, the initial results of that pilot program lead by the University of Rochester and Teledyne Imaging Systems were encouraging \citep{Bacon.2003a,Bacon.2004a,Bacon.2004b,Bacon.2010a}. 

Motivated by the possibility of discovering NEOs in thermal IR wavelengths with WISE but aware of its limitations in terms of lifetime and field of view, our group revisited the space-based NEO survey designs.  The major advance that resulted in WISE's improved sensitivity and spatial resolution over IRAS was the increase in its focal plane formats to 1024x1024 from 62 pixels, despite WISE's smaller primary mirror (40 cm vs. 60 cm).  Moreover, the focal planes of the \emph{Spitzer Space Telescope} \citep{Werner.2004a} were predicted to equilibrate at $\sim$30 K after the liquid helium was exhausted, a prediction that was borne out when the focal planes stabilized at 29 K after the cryogen ran out in 2008 \citep{Storrie-Lombardi.2012a}.  This prediction suggested that if 10 $\mu$m HgCdTe arrays could be enlarged to a 1024x1024 format and could achieve high operability at $\sim$35 K, it would be possible to cool the detectors and telescope purely passively.  An orbit that maintains enough separation from the Earth to minimize its heat load while still maintaining a nearly constant distance, thus supporting high-bandwidth communications, can be found at either of the Earth-Sun L1 or L2 Lagrange points.  While WISE is constrained by its orbit and sunshade to viewing a narrow strip close to 90$^{\circ}$ solar elongation, a mission at L1 with a taller sunshade can point much closer to the Sun.

The resulting mission concept, called the Near-Earth Object Camera (NEOCam) was first proposed to NASA's Discovery program in 2005 and again in 2010.  In 2010, the project was awarded technology development funding to mature the long-wavelength IR HgCdTe arrays.  The results of that effort have produced new 10 $\mu$m-cutoff 1024$^{2}$ arrays with very high pixel operability at 35-40 K \citep{McMurtry.2013a}.  The mission design allows for long lifetime, high data transfer rates supporting downlink of full-frame images, and the ability to view large swaths of Earth's orbit instantaneously.   By enabling full-frame downlinks, standard astronomical data processing techniques for extracting sources and producing accurately calibrated astrometry and photometry can be employed.  Many objects are detected at or near the detection threshold, so the ability to extract faint sources and distinguish them from noise and artifacts is essential.  With NEOWISE, we have demonstrated the ability to extract sources down to signal-to-noise (SNR) of 4.5 with extremely high reliability and completeness \citep{Cutri.2012a, Mainzer.2011a}.  Similar science data processing techniques can be used for NEOCam.

Alternative orbits for carrying out an advanced NEO survey have been considered.  For example, \citet{Tedesco.2000a} and \citet{Cellino.2004a} considered a Venus-like orbit on the grounds that the shorter synodic period with respect to Earth might afford more observations of NEOs with particularly Earth-like orbits.  In theory, such an orbit might allow NEOs to be discovered more rapidly than a system operating at L1.  However, a Venus-like orbit takes a spacecraft $\sim$30--170 times further from the Earth relative to L1.  

At present, Ka-band \citep[the current state-of-the-art for spacecraft telecommunications systems, as demonstrated by the Kepler mission;][]{Koch.2004a} can support $\sim$150 Mbps downlink from L1; however, a spacecraft at Venus incurs a factor of $\sim$900--30,000 drop in data rate relative to L1.  Methods to radically increase data rates such as optical spacecraft communications systems have yet to be adopted for routine use by flight projects.  

The significant decrease in data rate from Venus means that full-frame images cannot be downlinked without lossy compression, increasing the difficulty of discovering NEOs at or near the sensitivity limit.  While the ``windowing" method was used on the Kepler mission to select only small groups of pixels around targets of interest, this technique requires a priori knowledge of the targets' positions.  However, when searching for new NEOs, target locations are not known.  ``Windowing" to reduce data volume for an NEO survey therefore requires performing all source extractions on board the spacecraft.  Given limitations on memory and processor capacity, the ability to compute backgrounds and identify transient instrumental artifacts is greatly reduced.  Such techniques have not been demonstrated to produce calibrated sources at low SNR.  Moreover, survey projects such as the WISE and Two-Micron All-Sky Survey projects \citep[][]{Skrutskie.2006a} benefit substantially from the ability to reprocess science data using optimally estimated calibration products derived after the survey data were collected \citep{Cutri.2012a}.  If full-frame images are not collected, then reprocessing to apply best calibration products cannot be carried out, leaving many potential detections lost and ruling out the possibility of improved photometry and astrometry.  

A Venus-trailing IR telescope, at 0.7 AU from the Sun, is further stressed by the increased heat load on the thermal system, requiring active cooling for a cryogenic instrument whose focal planes must operate at $\sim$35 K.  The overall design and system complexity of a spacecraft that must now operate far from Earth rather than remaining relatively nearby is increased.    

To compare the asteroid detection capabilities of surveys operated from L1 and Venus-trailing orbit, we have carried out detailed simulations of these two scenarios.  Two simulation approaches were taken.  First, we performed a detailed simulation of a survey observing a \textbf{limited region of sky} over two years.  This approach includes the creation of spatially static sources (primarily stars and galaxies) based on extrapolation of WISE and \emph{Spitzer} data to NEOCam's predicted sensitivity limits as well as modification of the WISE Moving Object Processing System \citep[WMOPS;][]{Cutri.2012a,Mainzer.2011a} to accommodate the cadence required to discover NEOs without the need for ground-based follow-up (i.e. self-follow-up).  The simulated near-Earth asteroid (NEA; near-Earth comets were not included in this study) population is based upon the latest estimates from NEOWISE \citep{Mainzer.2011a,Mainzer.2012a}, with orbital elements taken from \citet{Bottke.2002a} as implemented by \citet{Grav.2011a}.  The second approach did not include static sources, but instead simulated a survey that covers \textbf{the entire viewable sky} over a six-year period.  Taken together, these two approaches allow new insights to be gained into the potential capabilities and optimization of a next-generation IR survey.

Both simulations follow this outline: 1) A frame-by-frame list of pointings for the survey is generated; 2) mission parameters are input (e.g. instrument FOV including detector gaps, sensitivity, slew time, downlink time, etc.); 3) a synthetic asteroid population that models expected numbers, orbital elements, and physical properties (e.g. diameter, albedo, etc.) is generated; 4) the predicted positions, infrared fluxes, and visible fluxes of each object in the population model are computed for each exposure in the simulated survey over six years using two thermal models; 5) objects that are detected the requisite number of times at the appropriate cadence are tallied. The population model is generated by randomly drawing orbital elements from the \citet{Bottke.2002a} model and assigning diameters and albedos randomly from the distributions given in \citet{Mainzer.2012a} for Atens, Apollos, and Amors.  The crucial difference between the limited-region and all-sky surveys (other than the area covered) is that in the limited-region simulation uses the modified WMOPS system to link detections together, allowing its efficacy to be evaluated.

\section{Survey Simulation Assumptions}
In this section, we describe the assumptions that were made for both the limited region and all-sky survey simulations.  

We assume that both L1 and Venus-trailing missions employ similar instruments with identical performance characteristics (i.e. telescope aperture size, detector performance, slew times, etc.).  Both instruments are assumed to cover the sky at the same rate, and both must obey the same restrictions on their cadence: that is, while each instrument has different viewing constraints owing to their different orbits, the sequence of observations in both position and time required to successfully discover NEOs is identical.  Even though there is a factor of $\sim$900-30,000 decrease in data rate for the Venus-trailing mission compared to L1 (necessitating  complex on-board data processing and destructive data compression that will degrade performance), we nevertheless assume that data can be extracted and processed to the same sensitivity limits and with the same astrometric and photometric precision as the L1 mission.  No penalty for lossy data compression was assumed.  If sources cannot be reliably extracted down to low SNR, then performance will be degraded accordingly.

\subsection{Instrument Performance Characteristics}
\begin{figure}[H]
 \figurenum{1}
 \includegraphics[width=3in]{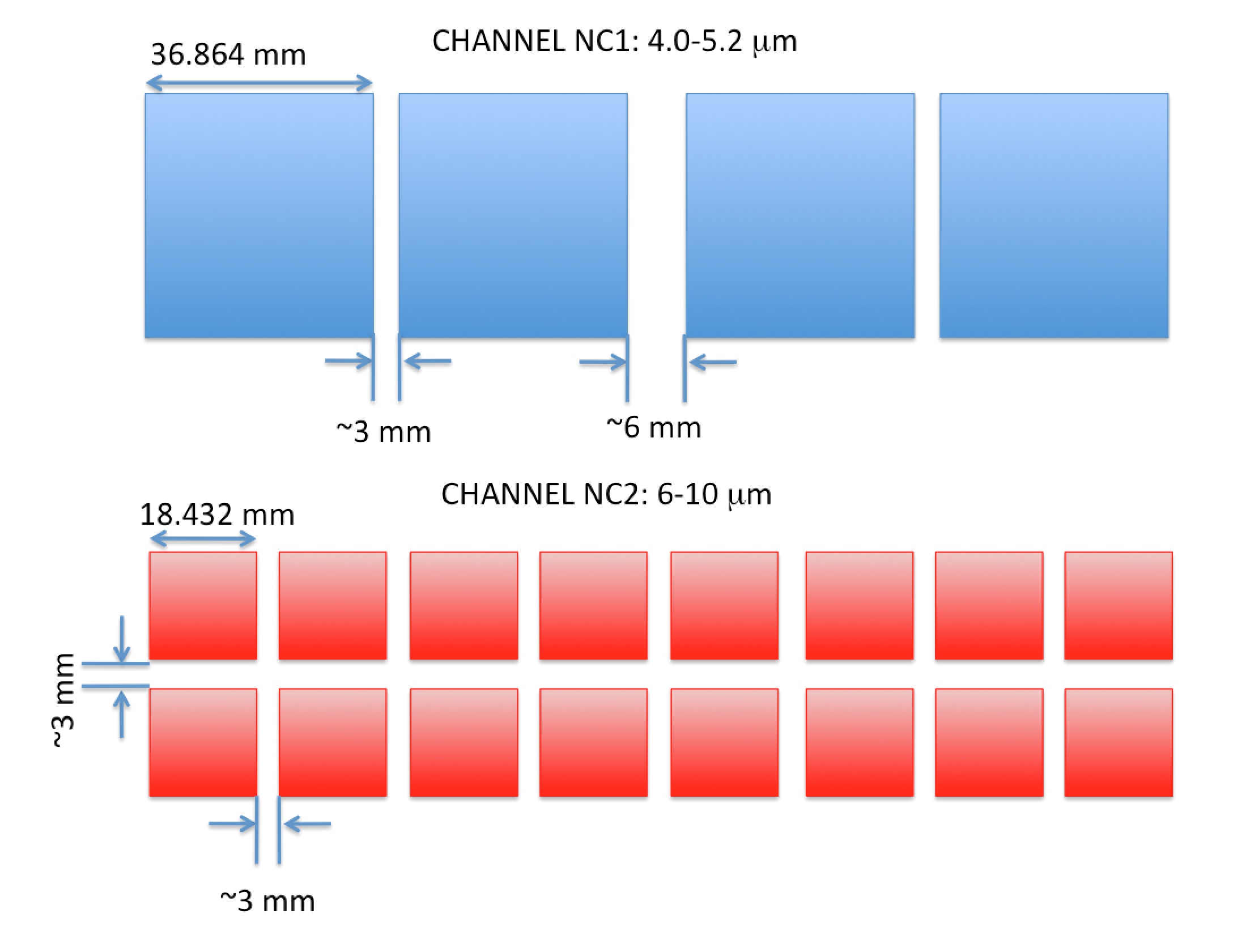}
 \caption{\label{fig:mosaic} Schematic showing the layout of detector arrays in mosaics for the two channels (4 - 5.2 $\mu$m, NC1, and 6 - 10 $\mu$m, NC2).  The NC1 channel consists of a mosaic of 2048 x 2048 HgCdTe arrays, while the NC2 arrays are 1024 x 1024 format. Both channels image the same field simultaneously using a beamsplitter.  The gaps between detector arrays were included in both all-sky and test region simulations.}
\end{figure}

\textbf{Instrument:} The requirements for both missions are summarized in Table 1. Both missions (L1 and Venus-trailing) employ a 0.5 m telescope with a 1.85$^\circ$ x 7.77$^\circ$ field of view.  Both telescopes are assumed to be diffraction-limited in their bands, and both require 30 sec to slew between fields and settle sufficiently for precision pointing.  Total integration time at each pointing is taken to be 180 sec in an effort to maximize sensitivity and minimize inefficiency lost to slew time.  Both missions employ a 2 x 8 mosaic of 1024 x 1024 10 $\mu$m HgCdTe detectors bonded to Teledyne Imaging System's HAWAII 1RG readout circuits similar to those described in \citet{McMurtry.2013a}, as well as a 1 x 4 mosaic of 5 $\mu$m HgCdTe detectors in a 2048$^{2}$ format.  The spacing between the arrays is shown in Figure \ref{fig:mosaic}.  Both L1 and Venus-trailing missions are assumed to have two bandpasses that simultaneously image the same region of the sky using a beamsplitter: 4 - 5.2 $\mu$m (referred to as NC1) and 6 - 10 $\mu$m (NC2).   

\begin{deluxetable}{lll}
\tabletypesize{\small}
\tablecolumns{3}
\tablecaption{Requirements for both an L1 and Venus-trailing survey.  Values for NC1 and NC2 channels are separated by ``;".  Read noise is specified for correlated double sampling (CDS).  *Since point source sensitivity varies with ecliptic latitude, longitude, and heliocentric distance, the point source sensitivity is specified near the midpoint of the viewing zones for each survey: ($\lambda$, $\beta$) = (0$^\circ$, 90$^\circ$) for L1, and (0$^\circ$, 180$^\circ$) at 0.7 AU heliocentric distance for the Venus-trailing survey. }
\tablehead{\colhead{Requirement} & \colhead{L1} &\colhead{Venus-trailing}  }
\startdata
Telescope aperture (m) & 0.5 & 0.5\\
Wavelengths ($\mu$m) & 4 - 5.2; 6 - 10 & 4 - 5.2; 6 - 10 \\
Field of view ($^\circ$) & 1.85 x 7.77 & 1.85 x 7.77 \\
Slew time (s) & 30 & 30 \\
Dwell time (s) & 180 & 180 \\
Minimum background current (electrons/s) & $\sim$300 & $\sim$300 \\
Detector format  & 1x4 2048$^{2}$ mosaic; & 1x4 2048$^{2}$ mosaic; \\
 & 2x8 1024$^{2}$ mosaic & 2x8 1024$^{2}$ mosaic \\
Detector type & HgCdTe & HgCdTe \\
Pixel size ($\mu$m) & 18; 18 & 18; 18 \\
Image FWHM (arcsec) & 3; 4 & 3; 4 \\
Dark current (electrons/s) & $<$5; $<$200 & $<$5; $<$200 \\
Read noise (CDS) & 15; $<$30 & 15; $<$30 \\
Quantum efficiency (\%) & $>$60; $>$60 & $>$60; $>$60 \\
Point Source Sensitivity* ($\mu$Jy, 5-$\sigma$) & 50; 150 & 50; 150 \\
Viewing zones (solar elongation) & $\pm$(45$^\circ$ --125$^\circ$) & 180$\pm$75$^\circ$ \\
Viewing zones (ecliptic latitude) & $\pm$41.9$^{\circ}$ & $\pm$41.9$^{\circ}$ \\

\enddata
\end{deluxetable}

The optical systems for both L1 and Venus-trailing missions are taken to be diffraction-limited at 4 $\mu$m.  The point spread function (PSF) full width at half-maximum (FWHM) for both missions is $\sim$4 arcsec in channel NC2, and the plate scale is set to 3 arcsec/pixel to maximize areal coverage at the expense of some loss of sampling.  The PSF will be undersampled for the NC1 channel. Astrometric uncertainty is taken to be $\sim$0.5 arcsec 1-$\sigma$.

\textbf{Zodiacal Background:} In the 6 - 10 $\mu$m band, an instrument's optical system and detectors must be cooled to achieve natural background-limited performance.  The dominant source of natural background is thermal emission from zodiacal dust.  In the 4 - 5.2 $\mu$m band, for an adequately cooled instrument, sensitivity is limited by a combination of dark current, read noise, and zodiacal background flux.  The zodiacal background is represented by a model that varies with ecliptic latitude and longitude based on the fluxes tabulated by \citet{Wright.1998a} and \citet{Gorjian.2000a}.  As described in \citet{Leinert.1998a}, the density of zodiacal dust increases as $1/r$, where $r$ is heliocentric distance. The requirements for detector dark current and read noise derive from the minimum zodiacal flux.   

\subsubsection{Orbits and Viewing Constraints}
\begin{figure}[H]
 \figurenum{2}
\includegraphics[width=6in]{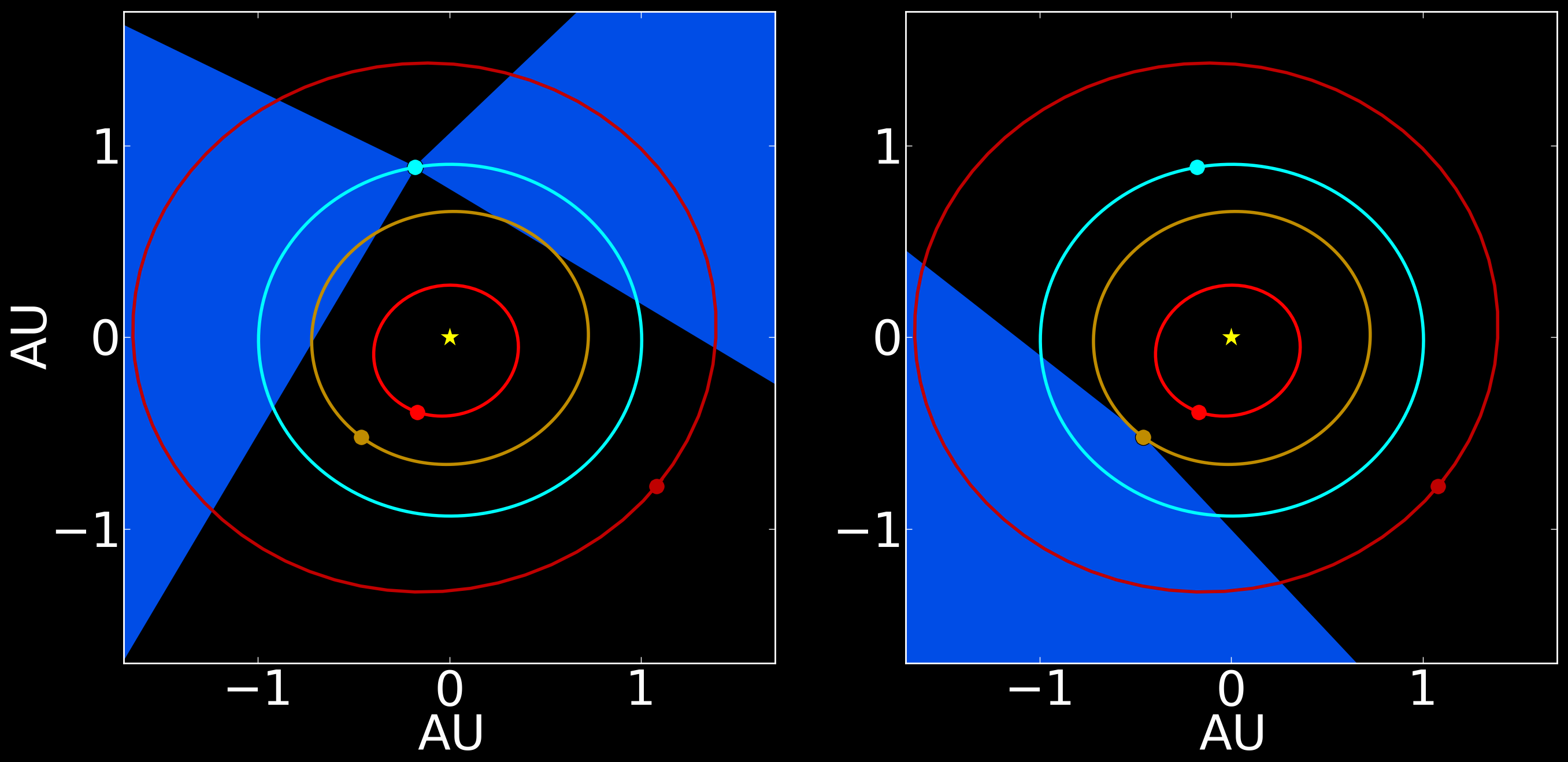}
 \caption{\label{fig:surveysim} Viewing constraints for both the Earth-Sun L1 Lagrange point (left) and the Venus-trailing (right) surveys as shaded areas.  The orbits and locations of Mercury, Venus, Earth and Mars are shown.}
\end{figure}

The asteroids likely to be the most hazardous and require the lowest amount of energy to reach from Earth  (i.e. those with the most Earth-like orbits) are most readily found by searching the space around the Earth's orbit \citep{Chesley.2004a}.  Therefore, it is desirable to design an optical system for a telescope located at the Earth-Sun L1 point to be able to look as close to the Sun as possible in order to subtend the largest fraction of Earth's orbit (Figure \ref{fig:surveysim}).  Similarly, the Venus-trailing mission must be designed to maximize the instantaneous viewing area.  

The L1 mission is shielded from the Sun sufficiently to allow it to point anywhere from 45 to 125$^\circ$ solar elongation in either direction.  IRAS could observe as close as 60$^\circ$ from the Sun \citep{Beichman.1988a}, and the Hubble Space Telescope has carried out observations as close as $\sim45^\circ$ solar elongation \citep{Na.1995a}.  The Venus-trailing mission's viewing zone is centered at opposition and can point up to 75$^\circ$ away in elongation.  Despite its larger instantaneous viewing zone area, the Venus-trailing mission cannot cover more ``fresh" sky per unit time than the L1 survey, and it must execute a survey cadence that supports NEO discovery.    

\subsection{Survey Cadence}
The Minor Planet Center (MPC), a division of the International Astronomical Union, arbitrates discoveries of new minor planet candidates.  In order for an object to be declared ``discovered," it must be observed over several days.  While this is sufficient to classify most objects as NEOs or more distant orbits, this observational arc is generally insufficient to allow objects to be recovered by observers at future apparitions using conventional wide-field imagers; such objects are often effectively lost.  For an object to be recoverable at its next apparition, it is generally necessary to extend its observational arc out to at least 10--25 days (preferably 60 to 90 days), depending on the observatory's astrometric accuracy.  That is, astrometric uncertainty can be no more than a few degrees off its prediction position after $\sim$3--10 years following its first detection.  It should be noted that observational arcs of a few weeks in duration are generally insufficient for formal hazard evaluation, and in general observational arcs spanning more than one opposition are required to reduce the orbital uncertainty sufficiently to allow location of the object within a few arc seconds.  However, even the shortest arcs described here, spanning $\sim$10 days, will allow linkage at subsequent return visits.  In essence, this is routinely accomplished by the MPC today as demonstrated by past recoveries and linkages of (719) Albert, 1937 UB (Hermes), and even 1954 XA.  

Unlike ground-based surveys, space telescopes at either L1 or Venus-trailing orbits can spend much of their time surveying the region of sky that is in the daytime sky for ground-based observers.  Therefore, the ability to perform ``self follow up" is essential because ground-based observers cannot be relied upon regularly for the short-term follow up required to determine orbits securely.  These space-based surveys are designed to reach dimmer limiting magnitudes than most existing ground-based follow-up stations (see Section 4.1 and Figure \ref{fig:visible}).  Future surveys must employ a cadence that ensures detection, follow-up, and accurate orbit determination are built into routine operations.   

\begin{figure}[H]
\figurenum{3}
\includegraphics[width=6in]{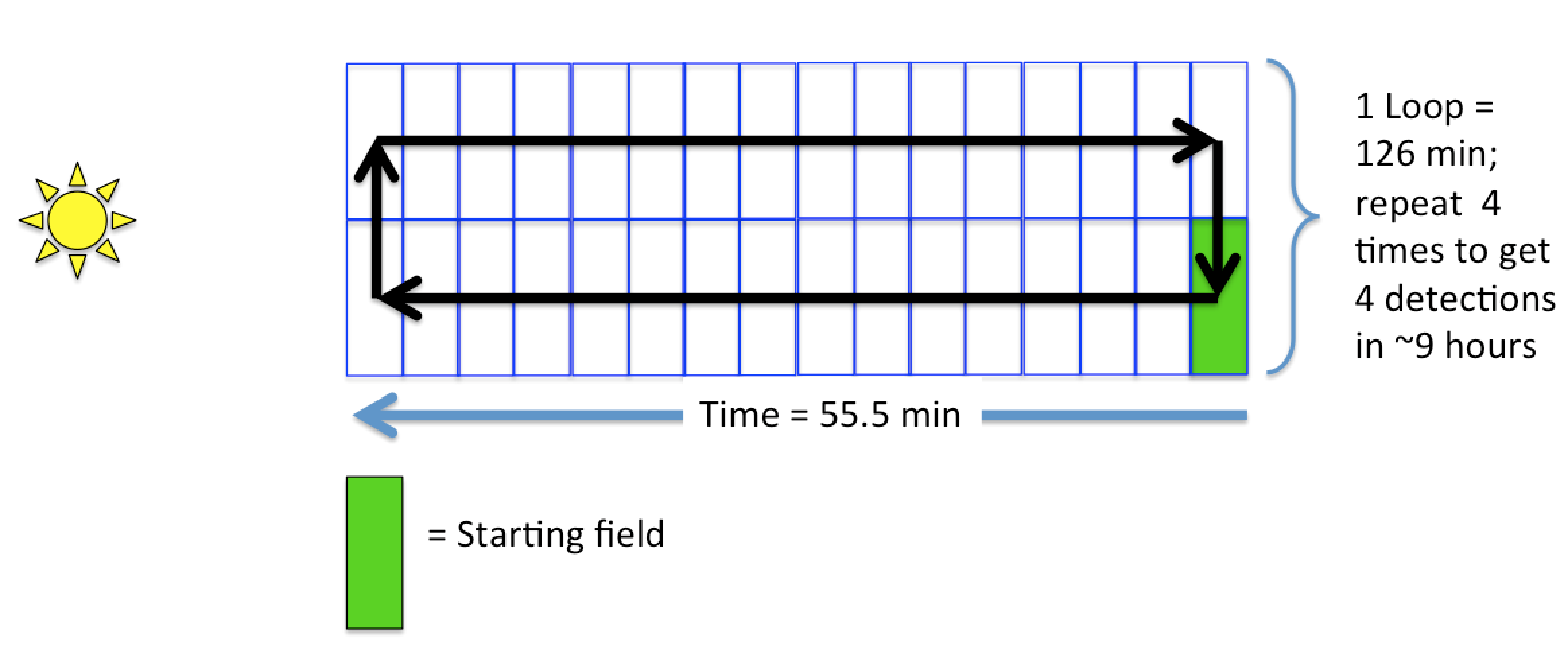}
\caption{\label{fig:cadence1} The basic element of the survey cadence is a ``quad," which consists of four observations taken over $\sim$9 hours.  The green rectangle indicates the starting position.}
\end{figure}

\begin{figure}[H]
\figurenum{4}
\includegraphics[width=6in]{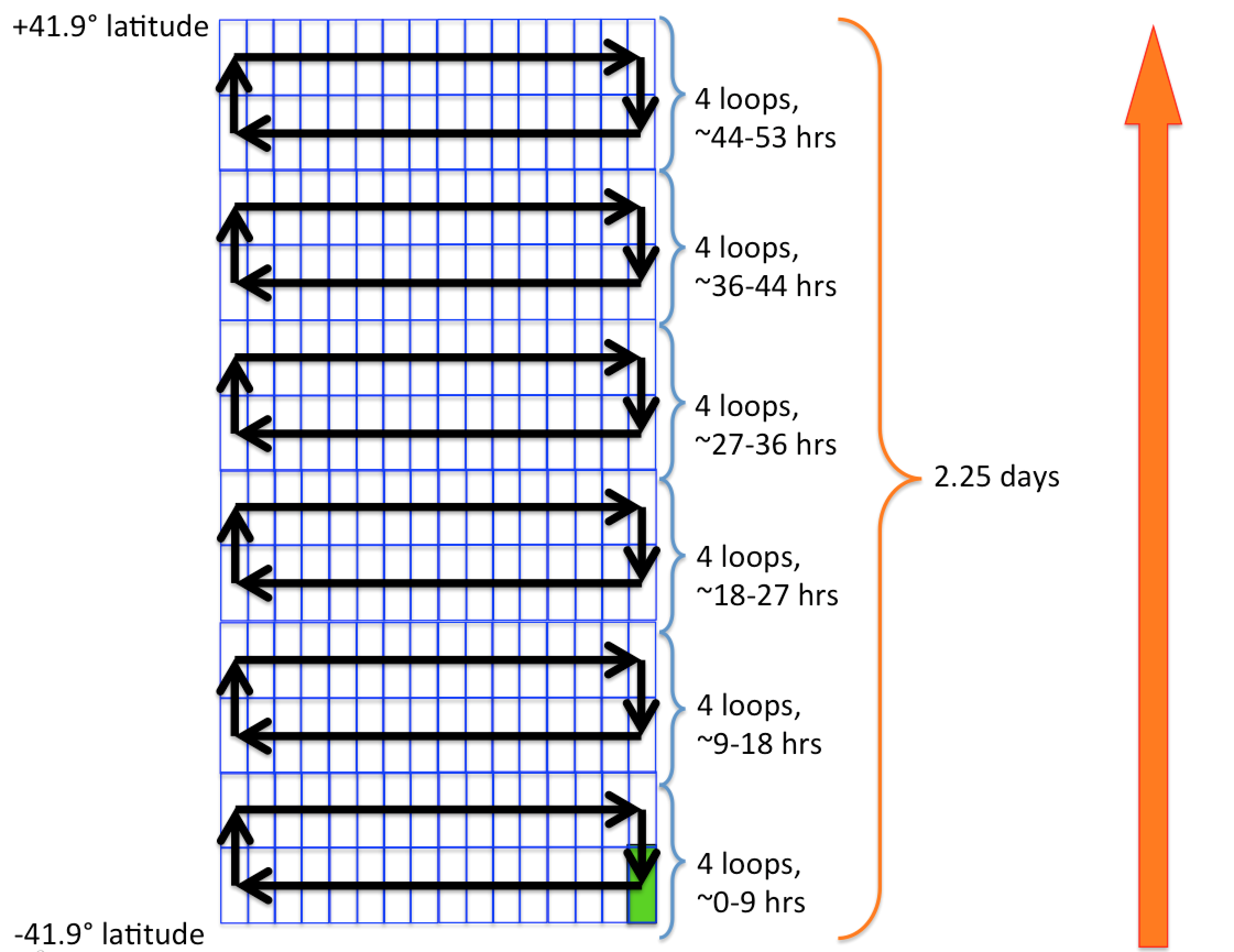}
\caption{\label{fig:cadence2} The loop pattern is repeated, increasing in ecliptic latitude, over $\sim$2.25 days. The green rectangle indicates the starting position.}
\end{figure}

\begin{figure}[H]
\figurenum{5}
\includegraphics[width=6in]{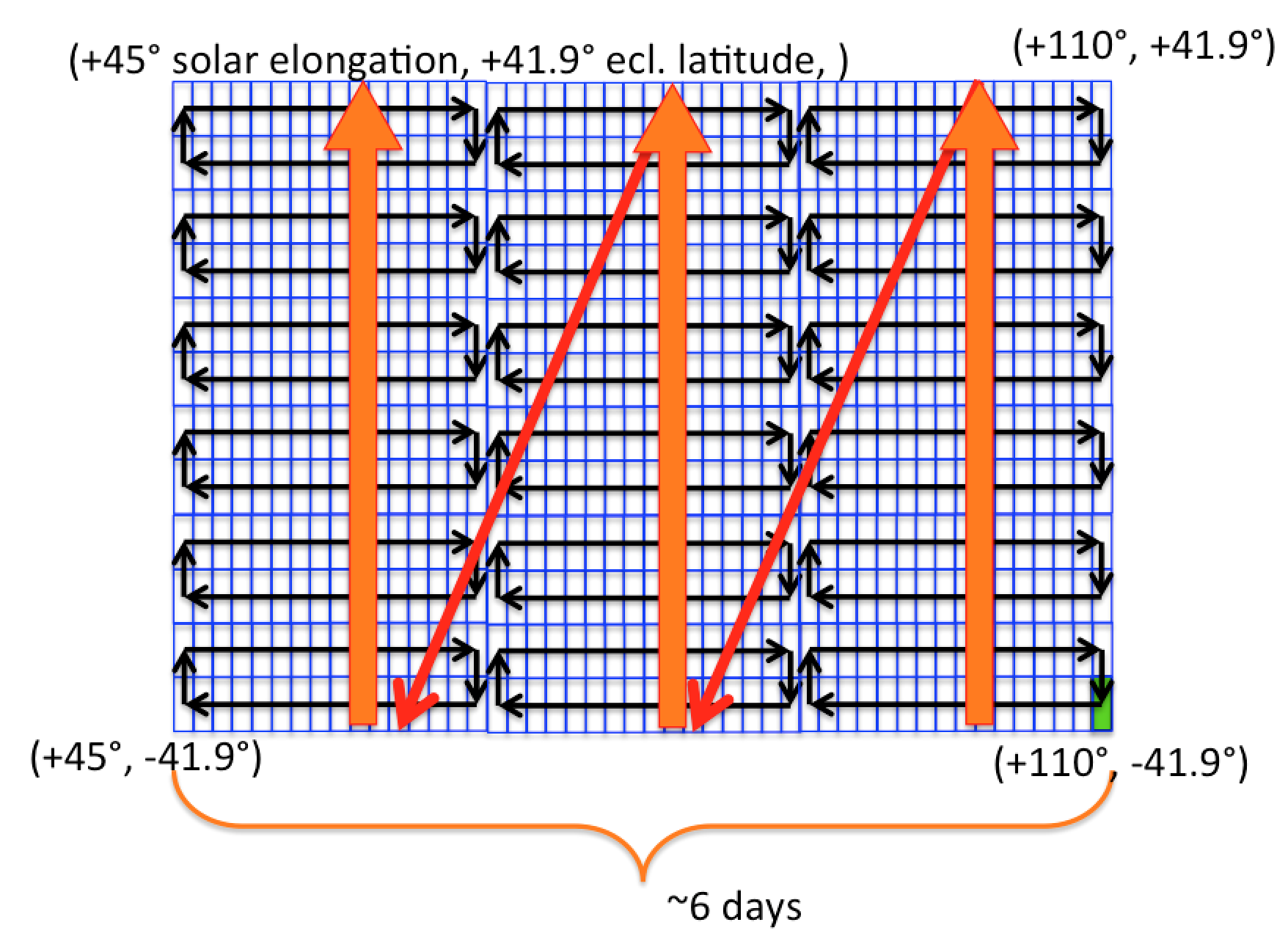}
\caption{\label{fig:cadence3} The survey pattern over $\sim$6 days for the L1 survey.  Green indicates the starting field.}
\end{figure}

\begin{figure}[H]
\figurenum{6}
\includegraphics[width=6in]{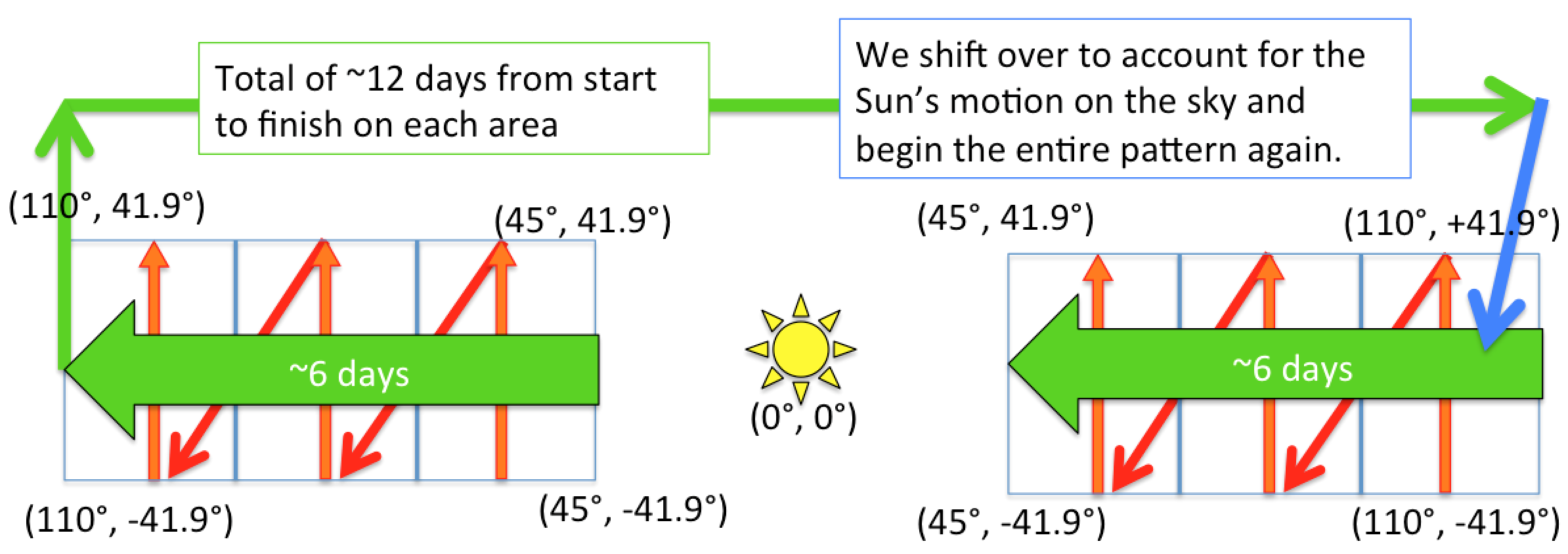}
\caption{\label{fig:cadence4} The entire cadence takes $\sim$22 days to execute, switching back and forth on either side of the Sun for the L1 survey, and proceeding in a continuous sweep for the Venus-trailing survey.  }
\end{figure}

Experience at the MPC has shown that a $>$20 day arc with approximately a dozen or more observations spaced  over time is required to fit an orbit with sufficiently low astrometric uncertainty to allow next-apparition recovery.  The cadence we tested for both the L1 and Venus-trailing surveys is illustrated in Figures 3--6 (spherical geometry has been neglected for simplicity).  The sequence begins with a set of four images taken $\sim$1-1.5 hours apart (a ``quad"), spanning $\sim$9 hours (Figure \ref{fig:cadence1}).  This cadence has been demonstrated by ground-based surveys such as Catalina to be robust against false linkages.  The survey starts at the maximum solar elongation and lowest ecliptic longitude.  Each step is taken along the shorter direction of the detector mosaic to minimize slew time.  Next, this loop pattern is repeated an additional five times, each at increased ecliptic latitude up to the maximum, over the course of $\sim$2.25 days (Figure \ref{fig:cadence2}).  For the L1 survey, this block of fields is surveyed over $\sim$6 days on one side of the Sun (Figure \ref{fig:cadence3}).  Next, an identical pattern is surveyed for the following $\sim$6 days on the other side of the Sun, after which point, the survey returns to the original position and repeats the sequence a total of 11 days later (Figure \ref{fig:cadence4}).  Both the L1 and Venus-trailing surveys cover sky at the same rate, and both result in $\sim$22 day observational arcs for a large fraction of the NEOs observed.

\subsection{Synthetic Asteroid Population Models}
For both all-sky and test region simulations, we considered NEAs as small as 140 m in effective spherical diameter.  The population models included Atens (NEAs with aphelia $Q>$0.983 AU and semi-major axes $a<$1.0 AU), Apollos (NEAs with $a>$1.0 AU and perihelia $q<$1.017 AU), Amors (NEAs with $a>$1.0 and $q>$1.017 AU), and interior-to-Earth objects (IEOs, or Atiras; NEAs with $a<$1.0 AU and $Q<$0.983 AU).  For the all-sky simulation, the synthetic solar system orbital element models of both \citet{Grav.2011a} and \citet{Greenstreet.2013a} were used to generate 25 randomly drawn populations of Atens, Apollos, and Amors.  By running 25 Monte Carlo trials, we average over variations in orbital elements and physical properties in randomly generated populations.  \citet{Greenstreet.2012a} was used to generate the IEO populations' orbital elements, as these were not included in the \citet{Grav.2011a} model, which is based on the orbital element distribution of \citet{Bottke.2002a}.  The numbers of Atens, Apollos, and Amors for the 25 synthetic populations were chosen according to the predicted total numbers found by \citet{Mainzer.2011b, Mainzer.2012b}, for a total of $\sim$13,200$\pm$1900 NEAs $>$140 m in each synthetic population.  For the test region simulation, only a single simulated population was generated, containing 12,700 NEAs $>$140 m.  The simulation does not at present consider near-Earth comets or long-period comets that enter near-Earth space.

Minimum orbital intersection distances \citep[MOIDs;][]{Bowell.1993a} were computed for each synthetic object using the same methods that are employed by the JPL Horizons system \citep{Ostro.2004a}.  MOIDs were computed so that the synthetic NEOs could be separated into those that would be deemed potentially hazardous asteroids (PHAs), depending on their size. PHAs are formally defined as having Earth MOIDs $<$0.05 AU and having absolute $H<22$ mag, although \citet{Mainzer.2012b} suggest a diameter-based definition, since IR surveys do not directly sample $H$.  

\begin{figure}[H]
\figurenum{7}
\includegraphics[width=6in]{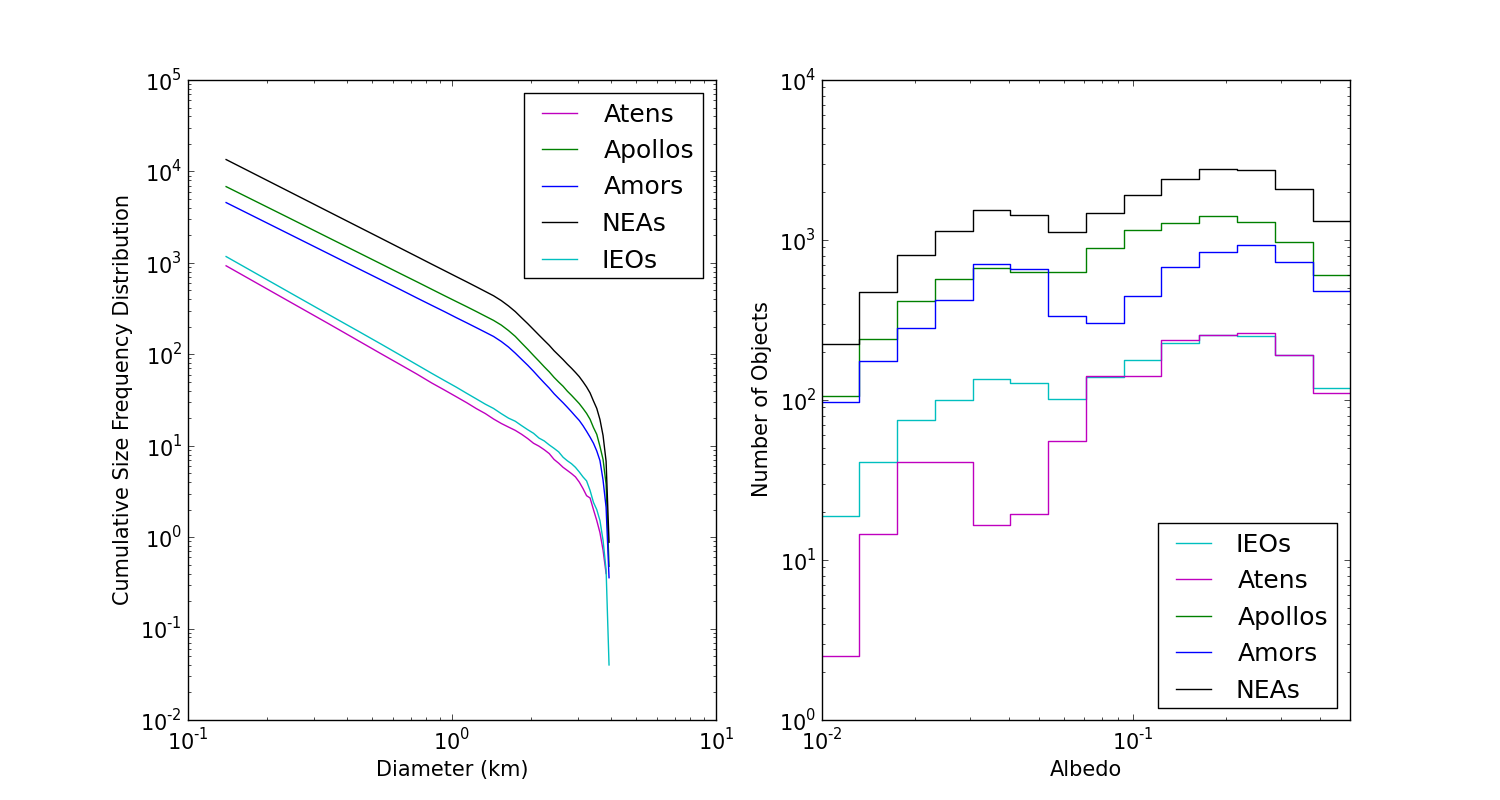}
\caption{\label{fig:sfd_albedo} Mean cumulative size frequency distributions (left) and albedo distributions (right) for the 25 synthetic populations of IEOs, Atens, Apollos, Amors, and all NEAs used in the all-sky survey simulation.}
\end{figure}

The physical properties for each object were randomly assigned according to the distributions given in \citet{Mainzer.2012b} for size, geometric albedo (\pv), beaming parameter $\eta$ employed by the Near-Earth Asteroid Model \citep[NEATM;][]{Harris.1998a} and infrared albedo (\pIR, the reflectivity at $\sim$3-5 $\mu$m) for the Atens, Apollos, and Amors, respectively.  \citet{Mainzer.2012b} found that the Atens, Apollos, and Amors each have different size and albedo distributions.  While the definitions of these three groups are somewhat arbitrary, the differences in physical properties probably reflect the differing source regions' properties for each group.  Since only 14 IEOs have been discovered to date, little is known about their total numbers or physical properties.  Populations of IEOs were generated using the \citet{Greenstreet.2013a} orbital element model.  Lacking information on the estimated total numbers of IEOs and their physical properties, the total numbers and size frequency distribution were assumed to be similar to the Atens; their albedo distribution was modeled based on the Apollos.  Figure \ref{fig:sfd_albedo} shows the median size and albedo distributions for all synthetic NEA subpopulations.   

We created 25 synthetic populations to lessen the chance that one individual randomly-generated set of asteroids could represent an outlier in terms of numbers of objects, orbital elements, or physical properties.  While the actual numbers, orbital element distributions, and size distributions are intended to be measured as a scientific goal of a next-generation survey,  this set of model populations allows comparison of two potential survey implementations (L1 and Venus-trailing).  

Fluxes were computed at NC1 and NC2 wavelengths for all objects at each time step in the simulated surveys using both the NEATM and the Fast Rotating Model \citep[FRM;][]{Lebofsky.1978a, Veeder.1989a, Lebofsky.1989a}.  Objects were modeled as faceted spheres \citep[c.f.][]{Kaasalainen.2004a} as described in \citet{Mainzer.2011b, Mainzer.2011c}.  The NEATM assumes a temperature distribution that decreases from the subsolar point to essentially zero at the terminator, whereas the FRM assumes a temperature distribution that is uniformly distributed in longitude.  While the NEATM assumes that the nightside of each asteroid contributes no flux, thermophysical models \citep[e.g.][]{Groussin.2011a, Delbo.2009a, Lebofsky.1989a} have shown that the assumption of zero nightside flux is not always good, particularly for observations at high phase angles where more of the nightside is seen.  Objects with high thermal conductivity will tend to have more uniform temperature distributions.  The NEATM also requires that the subsolar point is the hottest surface point.  Thermophysical models also show that with finite rotation and thermal inertia, the peak temperature is lowered and displaced from the subsolar point, altering the day side temperature distribution as well.  Small NEOs are known to have a higher fraction of fast rotators than larger objects \citep{Warner.2009a, Pravec.2008a, Pravec.2007a, Pravec.2000a}.  It is therefore useful to compare NEATM and FRM results as reasonable bounding cases.  The survey simulation results for the all-sky L1 and Venus-trailing surveys using both the NEATM and FRM are given in Section 4.1.

\section{Test Region Simulations}
In this section, we describe the results of simulations that were carried out for both Venus-trailing and L1 surveys on a limited test region spanning $\sim$260 square degrees populated with simulated NEAs, the full catalog of $\sim$600,000 known asteroids and comets, and static sources such as stars and galaxies.  The purpose of these detailed simulations was to test the efficacy of our moving object detection algorithms in the presence of potentially confusing transient noise and static sources.  The $\sim$260 square degree field is centered on ecliptic longitude/lattitude (+140$^{\circ}$, latitude 0.0$^{\circ}$), bounded by 129 to 151$^{\circ}$ longitude and -6 to +6$^{\circ}$ latitude.  

A list of field center positions and spacecraft ephemerides for the first two years of the Venus-trailing and L1 surveys were used as the basis for source list generation using the observing cadence described in Section 2.2.  A background ``static sky" was generated for each frame based on source count measurements from previous IR surveys as described below.  Fluxes for a population of  minor planets, including a synthetic NEA population and the full catalog of previously known Main Belt asteroids, Jovian Trojans, planets, etc., were computed for each frame and added to its source list.  Position-time sets for each candidate minor planet detection (``tracklets") were linked using a version of WMOPS that was adapted for the specific needs of the new cadence.  This version of the pipeline, dubbed the Experimental Moving Object Processing System (XMOPS), allowed us to estimate the tracklets' reliability and completeness as well as the frequency of cross-linkages between inertial sources and other asteroids.  The final results for the test region simulations are described in Section 3.4.

\subsection{Static Sky}
The static sky consists of astrophysical sources such as stars and galaxies that could potentially be confused with minor planets.  For our test region simulation, we constructed an artificial ``sky" using source catalogs from WISE and the \emph{Spitzer Space Telescope's} Infrared Array Camera \citep[IRAC; ][]{Fazio.2004a}, along with knowledge of the mid-IR sky \citep[e.g.][]{Jarrett.2011a}.  The survey simulation bandpass of NC1 is approximately the same as the WISE 4.6 $\mu$m (hereafter referred to as W2) and IRAC 4.5 $\mu$m (IRAC-2) channels, whereas the longer band, NC2, has a response that is closer to the IRAC 8 $\mu$m (IRAC-4) than the WISE 12 $\mu$m channel (W3).  Hence, we assume that IRAC-4 may be used to construct the NC2 sky.  However, since IRAC-4 images are only available over limited regions of the sky, such as the Spitzer Wide-area Extragalactic Survey \citep[SWIRE;][]{Lonsdale.2003a}, we use the all-sky W2 and W3 densities to predict the IRAC-4 and hence NC2 sky.   

\begin{figure}[H]
 \figurenum{8}
\plottwo{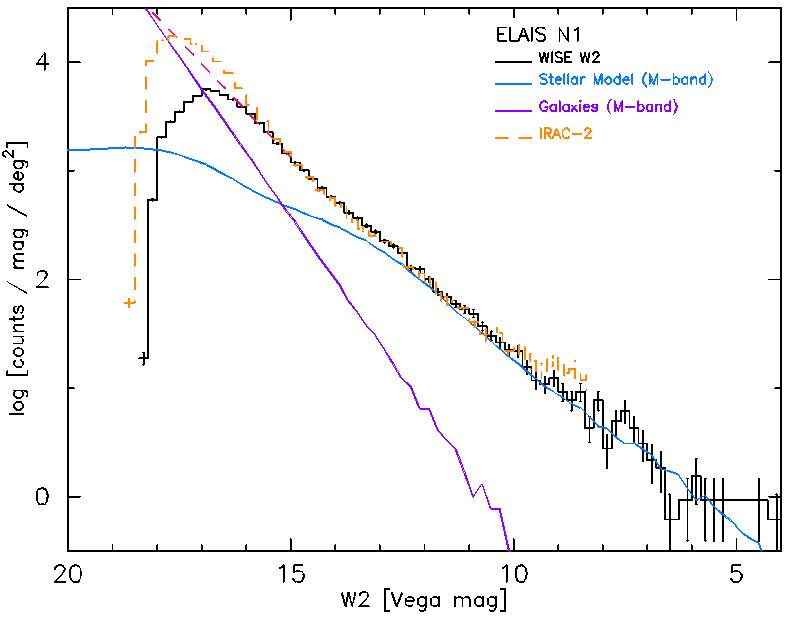}{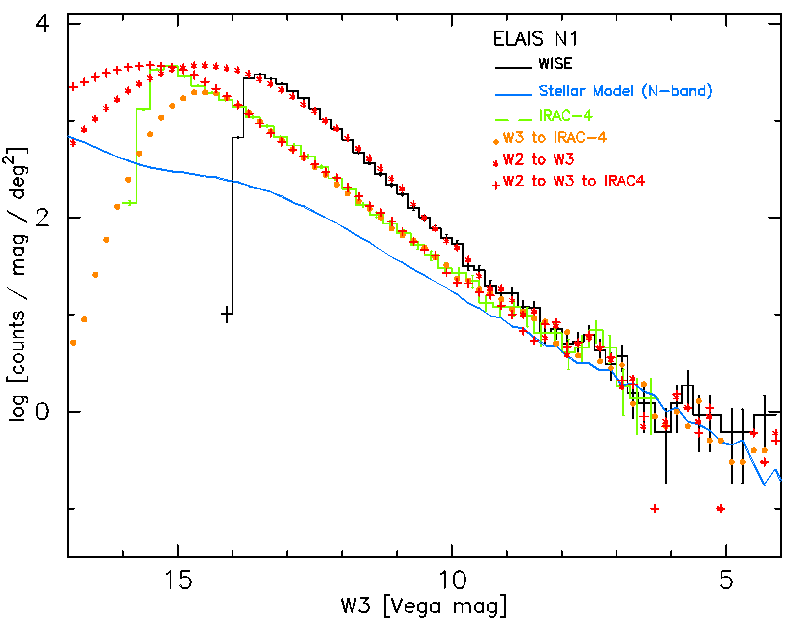}
 \caption{\label{fig:elais_sourcecounts} Left: Expected source counts for foreground stars in the Milky Way (blue solid line) and background galaxies (purple solid line) based on W2 (4.6 $\mu$m) and IRAC-2 (4.5 $\mu$m) source counts from the ELIAS-N1 region; both W2 (black solid line) and IRAC-2 (orange dashed line) are very similar to NC1 (4.6 $\mu$m).  At the faint end, the red dashed line shows the extrapolation to fainter magnitudes.  Right: Expected source counts for stars (blue solid line) and source counts from IRAC-4 (8 $\mu$m; green dashed line) and W3 (12 $\mu$m; black solid line).  The red and orange points show the predicted total source counts for sources in the IRAC-4 band (taken to be a proxy for NC2) based on extrapolation from W2, W3 and IRAC-4. } 
\end{figure}

We did a series of tests to determine the best method for extrapolating the static sky in the NC1 and NC2 bandpasses.  We used the SWIRE and WISE catalogs of the European Large Area Infrared Space Observatory Survey North region \citep[ELAIS-N1;][]{Rowan-Robinson.1999a} to explore the relationship between IRAC-2 and W2, and IRAC-4 to W2 and W3. Figure \ref{fig:elais_sourcecounts} (left) shows the source counts in W2 and IRAC-2 from the ELAIS-N1 region.  The plot includes the expected source counts for foreground stars in the Milky Way, as well as background galaxies as determined in the near-infrared $M$ band.  W2 and IRAC-2 agree reasonably well for W2 $<$ 16th mag (68 $\mu$Jy), but at fainter magnitudes WISE becomes incomplete and is diminished relative to IRAC-2.  The figure also shows the extrapolation to fainter magnitudes, a necessary step because WISE is not deep enough to reach the predicted sensitivity level in the NC1 band.  Although the fit to the WISE counts appears reasonable, the extrapolated source counts slightly underestimate the IRAC-2 counts at fainter flux levels (IRAC-2 appears to be incomplete at about $\sim$20 $\mu$Jy).  The discrepancy between the W2 and IRAC-2 at the faint levels likely arises from both incompleteness (WISE) and the Eddington bias, which causes fluxes to be overestimated at the faint end (IRAC-2).

\begin{figure}[H]
 \figurenum{9}
\plottwo{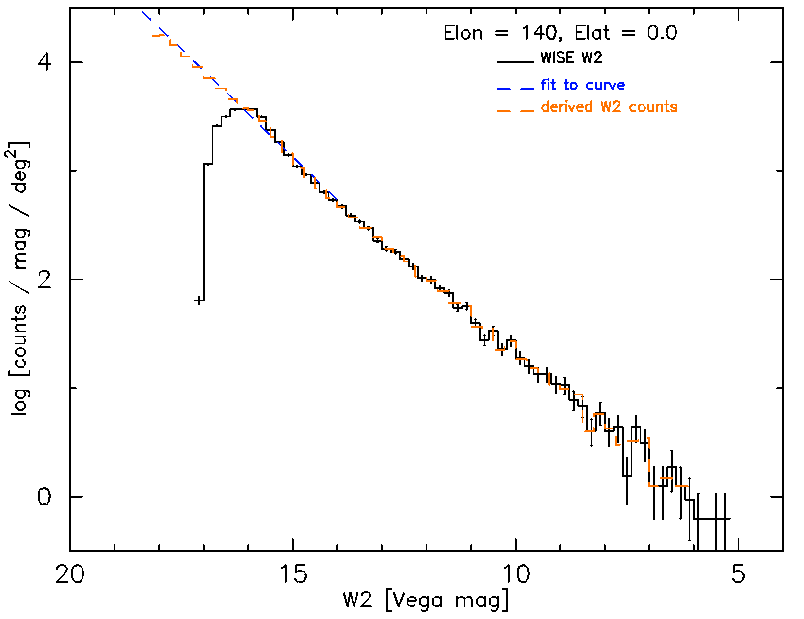}{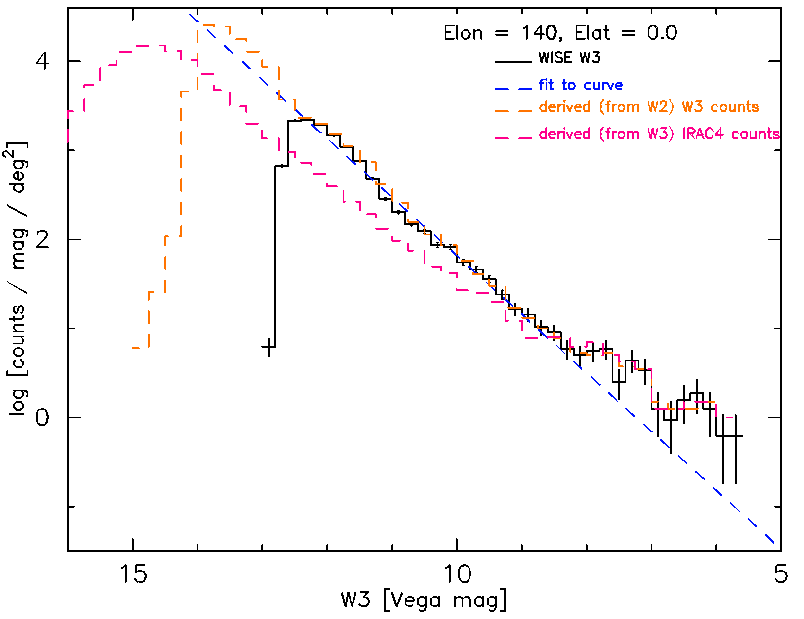}
 \caption{\label{fig:sourcecounts} Left: A fit to the W2 source counts (blue dashed and black solid lines, respectively) produces the extrapolated NC1 source counts (orange dashed line).  The W2 counts become incomplete at $\sim$16th magnitude.  Right:  W2 and W3 are used to reconstruct the IRAC-4 counts (dashed magenta line), which are used as a proxy for the NC2 sky (see Figure \ref{fig:elais_sourcecounts} for method details). } 
\end{figure}

Figure \ref{fig:elais_sourcecounts} (right) shows the resultant W3 and IRAC-4 source counts for the ELAIS-N1 field.  The NC2 band (6 - 10 $\mu$m) is not closely aligned with that of W3 (which spans $\sim$7.5 - 16.5 $\mu$m), but it is reasonably close to that of IRAC-4 (8 $\mu$m).  Since we do not have IRAC-4 images for most of the sky, we must use the WISE W3 all-sky data to reconstruct the IRAC-4 sky.  We must also account for the source counts in W2 since the static sky is a catalog of bandmerged NC1 and NC2 sources.  We therefore use W2 to predict the W3 source counts, then use our knowledge of W3 to predict the IRAC-4 counts.  The band-to-band differences between W2 and W3 are large, so the relative fraction of stars (which are blue in color at these wavelengths) and galaxies (which are red) makes a significant difference; i.e. fields dominated by foreground Milky Way stars will have a different color than regions dominated by extragalactic sources.  Figure \ref{fig:sourcecounts} (right) shows that the W3 counts can be reasonably reproduced by using W2 as the primary seed, and then the IRAC-4 source counts can be predicted from W3 accordingly.  Although the experiment on this region of sky was successful, more work remains to be done to verify that this method is robust against large color differences caused by a different mix of stellar and extragalactic sources.  

\emph{Source counts for the test region simulation.}  Using the methods described above, we generated the static sky for the entire $\sim$260 square degree test region simulation. Figure \ref{fig:sourcecounts} (left) shows the resultant W2 counts.  The W2 counts become incomplete for W2 $>$16 mag (68 $\mu$Jy); fitting to the faint bins, the resulting fit may be used to reconstruct the W2 counts for all flux levels, notably those that are near the NC1 sensitivity limit.  To simulate the NC2 static sky, we use W2 and W3 to extrapolate the source counts for IRAC-4, which serves as a proxy for NC2 as shown in Figure \ref{fig:sourcecounts} (right).  The reconstructed W3 population agrees well with the observed W3 source counts down to the limit at which W3 becomes incomplete.  This good agreement gives confidence that the extrapolated IRAC-4/NC2 source counts represent the NC2 static sky accurately.  

As a final step in generating the static sky catalog for the test region, random right ascension and declination positions were assigned to ``new" sources that are fainter than the real W2 limit.  Position uncertainties were modeled as a function of intensity. 

\subsection{Moving Objects}
In order to create a population of non-NEAs that could create potentially confusing cross-links to NEA candidates, known asteroids were added to the XMOPS test frames.  The entire catalog of known asteroids was downloaded from the MPC, and objects with perihelia $>$1.3 AU, semi-major axes $<$34.19952, and eccentricity $<$1 were injected into the synthetic source lists.  A single synthetic NEA population consisting of 12,700 objects larger than 140 m was generated as described in Section 2.3 above. 

\subsection{Experimental Moving Object Processing System}
Source lists comprised of static sky backgrounds and synthetic MBA and NEAs were generated for each frame in the survey simulation that fell within the test region.  A two year interval was considered, from June 1, 2012 to June 1, 2014.  Asteroid fluxes in the NC1 and NC2 channels were generated using the NEATM.  Simulated position and magnitude were errors added to each source.    

WMOPS was modified slightly to operate on source lists generated according to the new survey cadence (three sets of four quads spanning $\sim$22 days).  The resulting system, XMOPS, was run on the two year's worth of simulated source extractions and compared to the known number of NEOs that appeared in each frame to compute an estimate of the efficiency with which tracklets could be found.  XMOPS provides the same core functionality as WMOPS, with the major differences being in the way in which regions are selected and grouped for processing.  The test region of sky was broken into smaller 3 x 3$^{\circ}$ subregions (``patches") that overlapped by $\sim$50\% or more.  Stepping along in 1.5$^{\circ}$ increments, each patch was checked to see if it had been imaged at least four times over a 24 hour interval.  If there were at least four coverages in 24 hours, the WMOPS routine that identifies inertially fixed sources with SNR$>$4 was run to remove sources that repeated at the same location from the source lists.  As described in Section 2.2, a minimum of four detections spanning $\sim$9 hours is required to form a ``quad".  A later threshold at SNR=6 was applied to the tracklet-detection input lists, and the tracklet efficiencies were used to inform the lower SNR samples (see Section 2.5.1).    

\begin{figure}[H]
 \figurenum{10}
\includegraphics[width=3in]{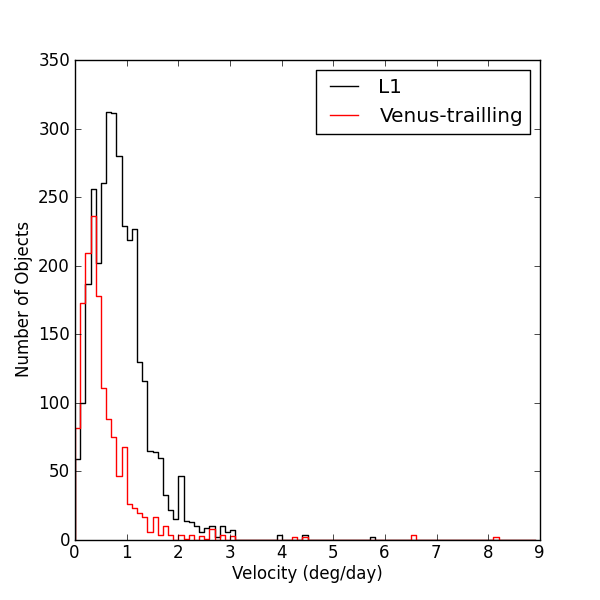}
 \caption{\label{fig:velocity} The average apparent on-sky velocity of NEAs in two years' worth of frames simulated in the $\sim$260 square degree test region searched with XMOPS.  NEAs detected by the Venus-trailing survey tend to have lower on-sky angular velocities (red line) because they are observed at greater heliocentric distances (this also makes them fainter).  Detections can only be linked if the apparent velocity is low enough to avoid significant trailing.} 
\end{figure}

\begin{figure}[H]
 \figurenum{11}
 \includegraphics[width=6in]{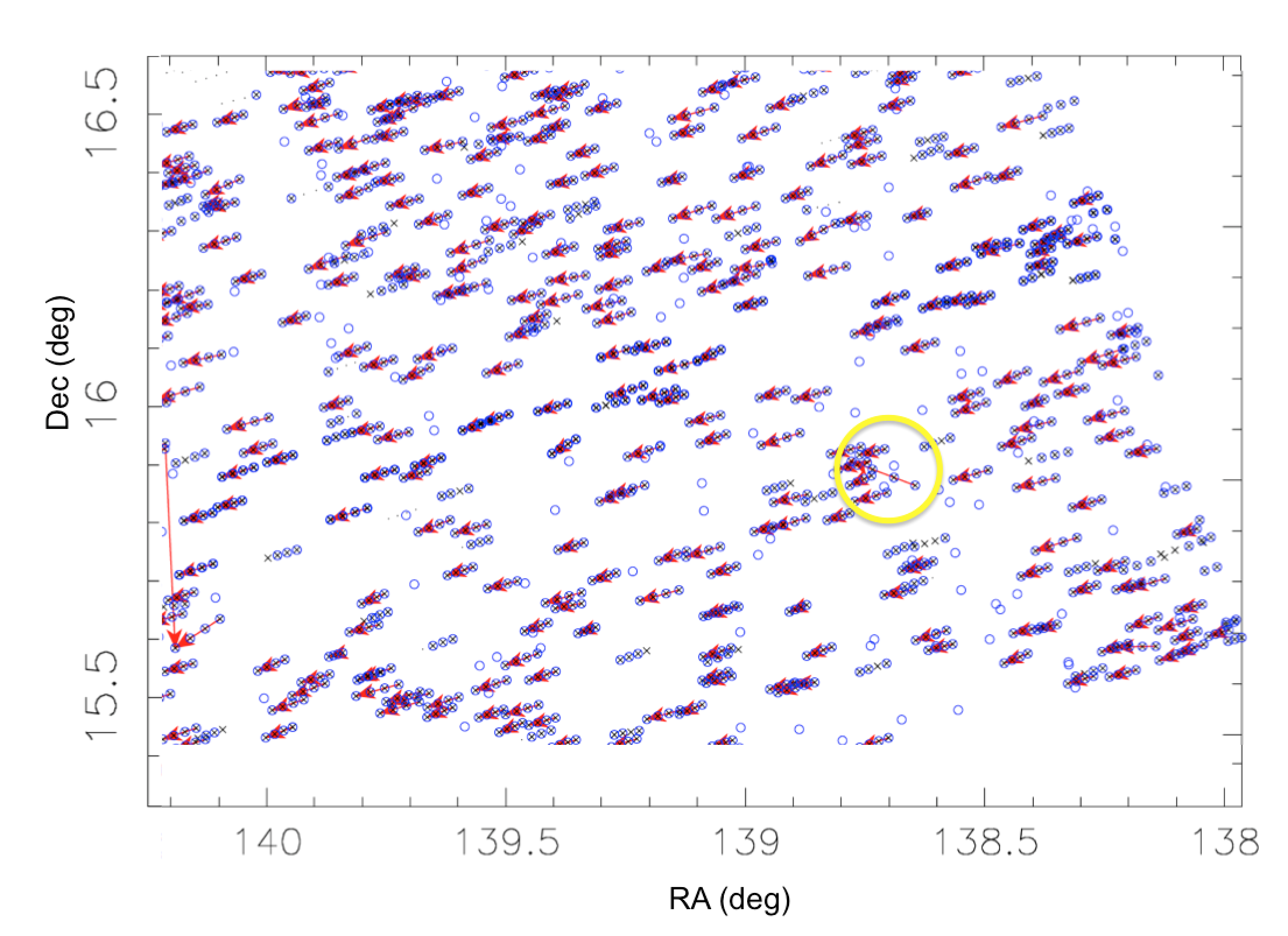}
 \caption{\label{fig:tracklets} A set of example candidate tracklets with four or more detections generated for one patch by XMOPS (red arrows indicate the start and end points of the tracklets).  Empty blue ``o" marks show the simulated detections that match the position, time, magnitude, and coverage criteria and were not identified as stationary sources.  Black ``x" marks show the expected, true known solar system objects that match the time/position window as well as the magnitude and coverage thresholds. An ``x" without an ``o" is a true, detectable solar system object that did not survive stationary object rejection, and an ``o" without an ``x" is a stationary source that survived stationary object rejection.  Any ``x" or ``o" that is not covered by a red arrow was not linked into a tracklet.  All but two of the red arrows represent Main Belt asteroids; the sole NEA in the frame is highlighted by the yellow circle.  An example of an erroneously cross-linked tracklet is shown by the nearly vertical red arrow on the extreme left. } 
\end{figure}

The \emph{FindTracklets} routine \citep{Kubica.2007a} was run to link pairs of candidate transient sources, similar to the methods employed by WMOPS \citep{Mainzer.2011a, Cutri.2012a}.  The \citet{Kubica.2007a} method uses hierarchical data structures called k-d trees to recursively partition the sources that could potentially be linked into smaller subsets, reducing the search time proportional to $\rho log \rho$ (where $\rho$ is the source sky-plane density) instead of $\rho^{2}$.  The \emph{CollapseTracklets} algorithm \citep{Myers.2008a} was used to link the detection pairs into candidate lists of moving object tracklets.  Minor adjustments were made to the parameters limiting how closely spaced in time the detections had to be as well as the maximum allowable velocity (Figure \ref{fig:velocity}).  Since WMOPS relied upon an ``eyes-on" human quality assurance step in which new candidate object tracklets were visually inspected, but a similar system was not implemented in this simulation, the maximum allowable velocity was reduced slightly to increase the reliability of tracklet linkages.  An example of candidate tracklets generated by XMOPS is shown in Figure \ref{fig:tracklets}.  The incidence of cross-linkages between two different asteroids and stationary sources that were not correctly identified by the static sky rejection algorithms were evaluated by comparing XMOPS outputs to the list of true tracklets.

\subsection{Results}
For the L1 mission, there were 1,123 NEAs that appeared with the minimum number of coverages ($\geq$4), but only 341 of those appeared above the SNR=6 cutoff.  Of these objects, 313 were successfully linked into one or more tracklets, yielding a linking success rate of 92\%.  The 112 NEAs that were detected in two or more tracklets represented 87.5\% of the 128 NEAs detected above the SNR=3 threshold.  These linking success rates are nearly identical to the results for the Venus-trailing survey, which linked 160 out of 170 NEOs above the SNR=6 cutoff into one or more tracklets (94\%), and 47 out of 55 NEAs into two or more tracklets (85\%).  Only 385 NEAs appeared above the coverage threshold for the Venus-trailing survey during the two year simulation; the number is considerably decreased relative to the L1 mission because the observatory is circulating at Venus' orbital period rather than Earth's, and the XMOPS portion of the simulation does not cover the entire sky.     

The nearly identical tracklet linking rates for both surveys suggests that there is no particular advantage or disadvantage to either orbit insofar as the ability to link tracklets is concerned, assuming that sources can be reliably extracted to low SNR.  A limitation of the XMOPS simulation at this point is that it did not consider the impact of spurious detections of artifacts and noise on the number of possible combinations that can create a tracklet.  If spurious detections cannot be reliably identified and excluded from the list of transient sources, they will overwhelm the pipeline with false cross-linkages.  The ability to correctly identify and exclude transient artifacts such as stray light, cosmic rays, and noisy pixels is essential.  

\section{All-Sky Simulation}
A second set of simulations was created to estimate the theoretical best performance that an advanced survey could achieve over the entire viewable sky for a six year survey for both L1 and Venus-trailing missions.  These models did not make use of XMOPS in the presence of a static sky.  Instead, we produced a list of every field pointing of the six year surveys and compared this to the positions of populations of simulated objects.  We generated synthetic orbital elements for each simulated object, and then numerically integrated the positions of these objects forward over the duration of the survey under the gravitational influence of the Sun, the eight planets, and the Moon with one hour timesteps.  We used the SWIFT numerical integrator \citep{Levison.1994a} implementing the Bulirsch-Stoer integration method on JPL's high performance computing facilities to calculate the positions of each object over the course of our simulation.  Positions of each object over time were then compared with the date and coverage of each field pointing, and along with distance from the telescope used to determine detectability.

As described above, the natural sky background was assumed to vary according to the models and measurements of \citet{Wright.1998a} and \citet{Gorjian.2000a} as a function of ecliptic latitude and longitude.  The Venus-trailing survey's zodiacal background was assumed to increase as a function of heliocentric distance according to Figure 55 in \citet{Leinert.1998a}.  Both Venus-trailing and L1 surveys were assumed to have the properties given in Table 1 and were assumed to carry out the cadence described above.  The sensitivity of the instrument was computed over a grid of ecliptic latitudes and longitudes with 5$^{\circ}$ steps spanning the fields of regard of both Venus-trailing and L1 surveys ($\pm$41.9$^{\circ}$ latitude for both surveys; 45 -- 125$^{\circ}$ solar elongation on either side of the Sun for the L1 survey at $\sim$1 AU heliocentric distance, and $\pm$75$^{\circ}$ in elongation centered at opposition at 0.7 AU heliocentric distance for the Venus-trailing survey).  For each NEA's position in each pointing during the surveys, the sensitivity was determined by interpolation between grid points in the sensitivity model.  

In these simulations, an object was counted as detected if its predicted flux was above the SNR threshold and its velocity fell within an allowable range.  The upper and lower apparent velocity limits were determined by trailing losses and the instrument's astrometric accuracy for faint sources, respectively.   This survey simulation did not account for confusion with inertially fixed sources such as stars and galaxies, confusion with optical or electronic artifacts, or more distant asteroids that could potentially be confused with NEOs.  Therefore, these simulated surveys should be regarded as predicting the ``best-case" of an advanced IR survey appropriate to low-confusion regions of the sky.   

The all-sky survey simulations for both L1 and Venus-trailing missions were assumed to start on June 1, 2014 and end on June 1, 2020.  The all-sky survey assumes that sources are extracted down to SNR=5.  As described in Section 2.3, a total of 25 synthetic populations were generated.  Using 25 randomly drawn populations ensures that the results are less likely to be affected by outliers in orbital elements or physical properties.  The all-sky simulations were performed using NEA fluxes computed with the NEATM and fast rotating thermal models for both Venus-trailing and L1 surveys.   

Since an advanced survey will build upon the progress made by existing ground-based facilities, it was necessary to develop a method of estimating how many NEAs in a given size range would have already been discovered by these projects.  A synthetic survey of ground-based detections beginning in 1985 and proceeding in five year increments to the assumed end date of the advanced survey (June 1, 2020) was created.  Visible light fluxes were computed for each synthetic NEA between 1985 and 2014. Objects were declared ``found" by the visible light surveys if they appeared in the night sky with solar elongations $>120^{\circ}$, ecliptic latitudes $<40^{\circ}$, and apparent magnitudes brighter than specified limits for each survey epoch.   The limiting magnitude of each epoch was tuned to the sensitivity of each successive generation of NEO survey telescopes from 1985 to the present day.  The resulting $H$ magnitude distribution compares very well to the known population.  The all-sky simulations assume that the existing suite of ground-based surveys (e.g. Catalina and PanSTARRS) continue to operate throughout the duration of the L1 and Venus-trailing IR missions.  The comparison between a space-based IR mission and an enhanced ground-based survey will be the subject of future work.  Ground-based and space-based missions complement each other, since ground-based telescopes are particularly sensitive to Amors, whereas the space-based surveys detect a higher fraction of IEOs, Atens, and Apollos as described in Section 4.1 below.  While diameters can be determined using IR fluxes alone once a reliable orbit has been established, visible albedos can only be determined if objects are detected by both IR and visible light surveys, significantly enhancing the value of both datasets.

%

\subsection{Results}

\begin{figure}[H]
\figurenum{12}
\includegraphics[width=6in]{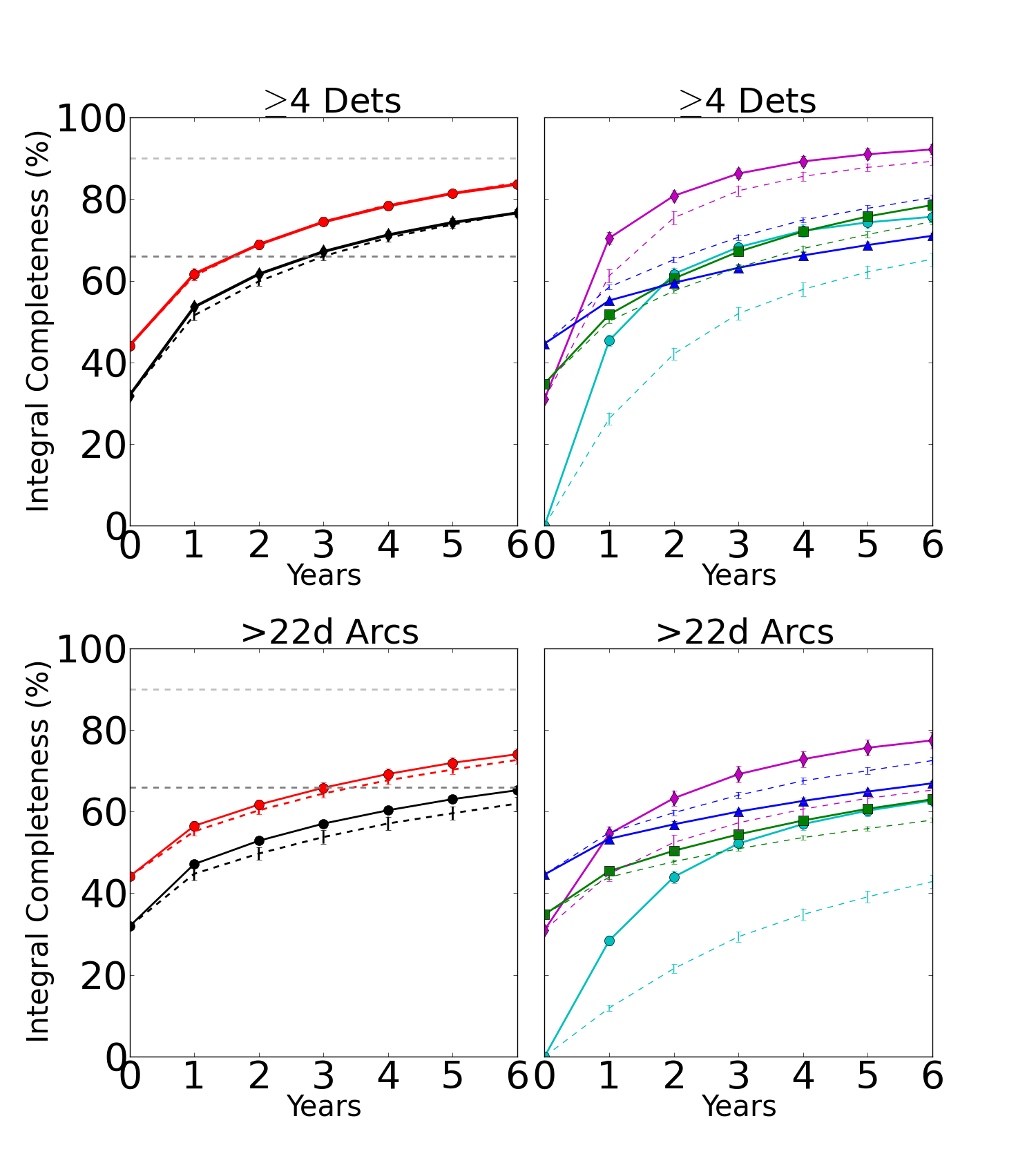}
\caption{\label{fig:NEATM} The integral survey completeness for populations of NEAs larger than 140 m in diameter versus time, with fluxes computed using the NEATM.  Solid lines represent the L1 survey; dashed lines represent the Venus-trailing survey. Dashed gray lines represent 2/3 and 90\% integral completeness limits, respectively.  Left plots: all NEAs (black lines) and PHAs (red lines), respectively.  Right plots: IEOs (cyan lines), Atens (magenta lines), Apollos (green lines), and Amors (blue lines).}
\end{figure}

\begin{figure}[H]
\figurenum{13}
\includegraphics[width=6in]{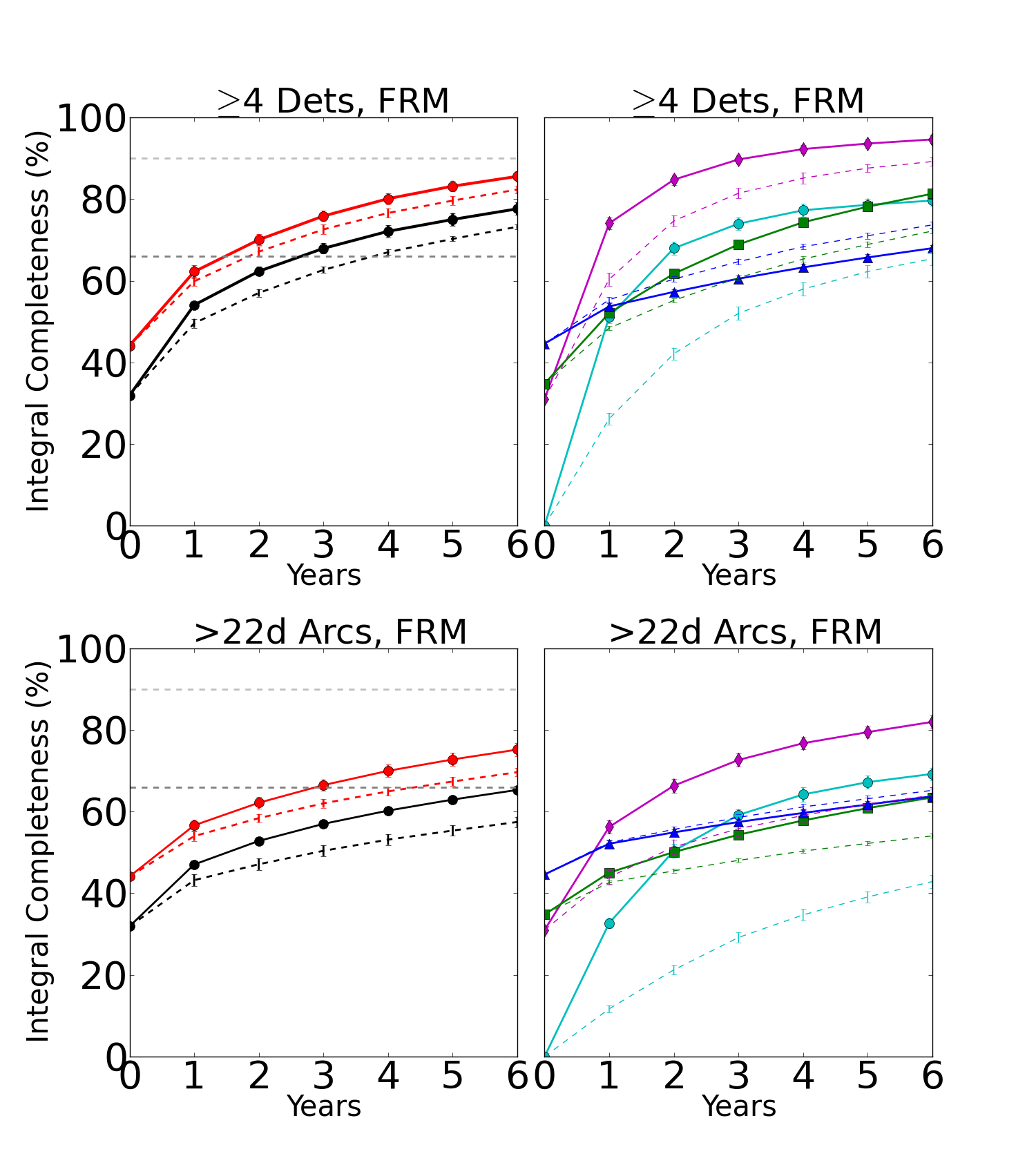}
\caption{\label{fig:FRM} The integral survey completeness for populations of NEAs larger than 140 m in diameter versus time, with fluxes computed using the FRM.  Color coding is the same as Figure \ref{fig:NEATM}.}
\end{figure}

Figures \ref{fig:NEATM} and \ref{fig:FRM} show the year-by-year integral survey completeness results for 140 m objects for NEATM and FRM, respectively.  Integral survey completeness is the total fraction of all objects greater than or equal to a certain size that have been detected. Estimates of the number of objects that were observed by the space-based surveys but were previously discovered by the ground-based surveys are included in the total.  The model of projected optical survey performance is a simple one and needs refinement, but it is adequate for the purpose of comparing the L1 and Venus-trailing surveys.  The top rows in both figures show the integral survey completeness for NEAs detected four times over $\sim$9 hours (an individual tracklet); the bottom rows show the integral survey completeness for NEAs detected with $>$22-day observational arcs.  At present, our simulations include only a relatively simplistic method for linking detections into tracklets across 22-day intervals in an attempt to replicate the function of the MPC.  We were not able to simulate more complex linkages, e.g., a ten-day observational arc followed by another set of ten observations carried out six months later.  Such a set of hypothetical observations is very likely to be linked by the MPC, resulting in a discovered object with a well-determined orbit.  Therefore, the $>$22-day arc results likely represent a conservative view of the surveys' capabilities.  The objects with four observations spanning $\sim$9 hours represent an upper limit, since objects are typically not designated as discovered until after they have received observations on two nights separated by $>$12 hours.  The one-day and 22-day results therefore represent bounding cases.

Figures \ref{fig:NEATM} and \ref{fig:FRM} show that the two surveys achieve similar integral survey completeness levels for NEAs $>$140 m.  The L1 survey outperforms the Venus-trailing survey for Atens, Apollos, and IEOs.  The Venus-trailing survey discovers more Amors because it observes around opposition.  However, Amors are significantly less likely to make close Earth approaches than Atens and Apollos, decreasing their potential hazard as well as their propensity to make suitable rendezvous targets.  The L1 survey slightly outperforms the Venus-trailing survey for PHAs in this size range.  The error bars in Figures \ref{fig:NEATM} and \ref{fig:FRM} were generated from the standard deviation among the 25 simulated populations.  The rather small error bars indicate that the estimates presented are not significantly affected by sampling error (a fact that could not have been reliably known without having done more than one simulation).  

\begin{figure}[H]
\figurenum{14}
\includegraphics[width=6in]{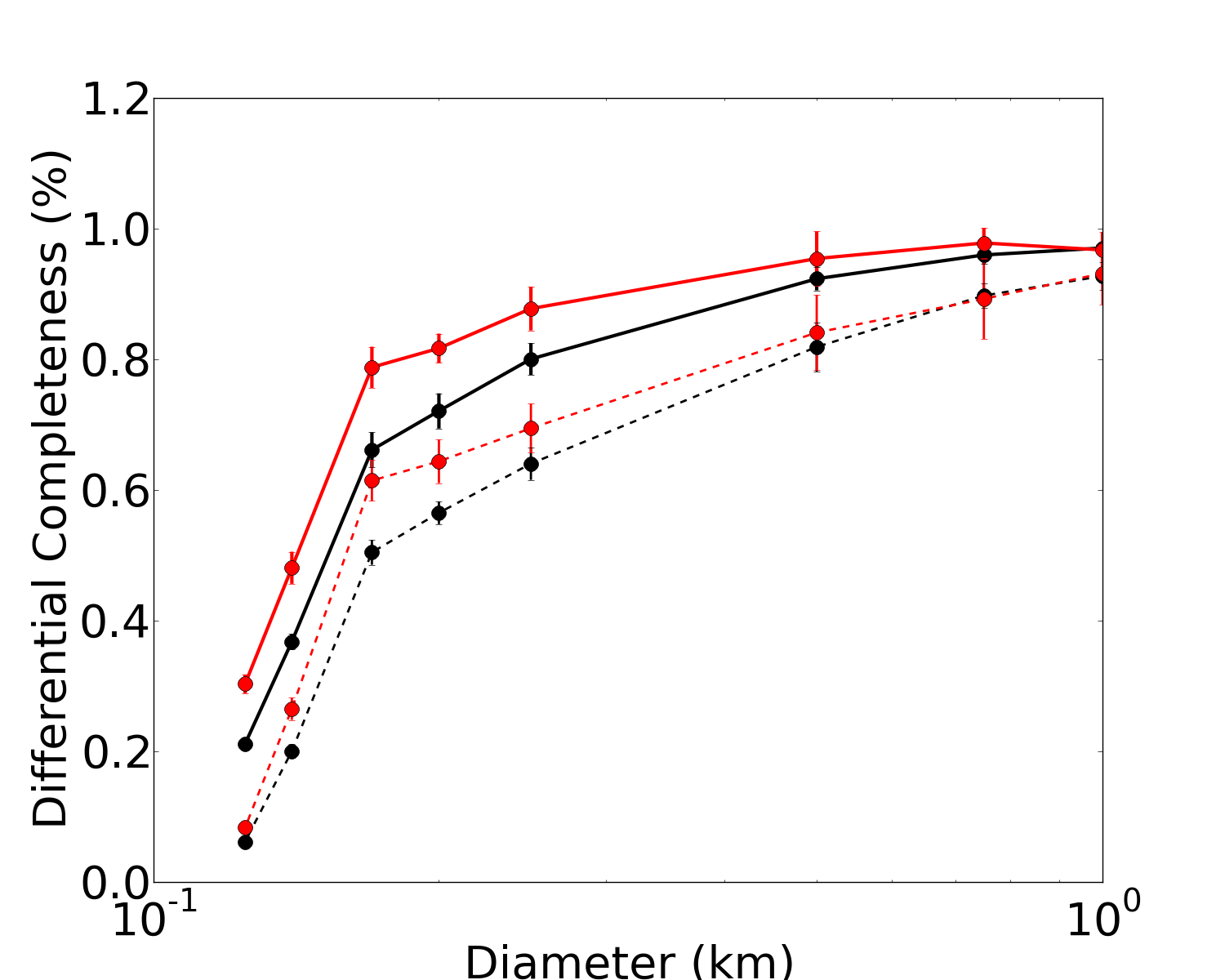}
\caption{\label{fig:diff_completeness} The differential survey completeness for populations of NEAs (black lines) and PHAs (red lines) larger than 140 m in diameter for a six year survey, with fluxes computed using the FRM.  Solid and dashed lines represent the L1 and Venus-trailing surveys, respectively.}
\end{figure}

Figure \ref{fig:diff_completeness} shows the NEA differential completeness for both surveys using the FRM after 6 years.  Differential completeness does not depend on an assumed size-frequency distribution.

\begin{figure}[H]
\figurenum{15}
\includegraphics[width=6in]{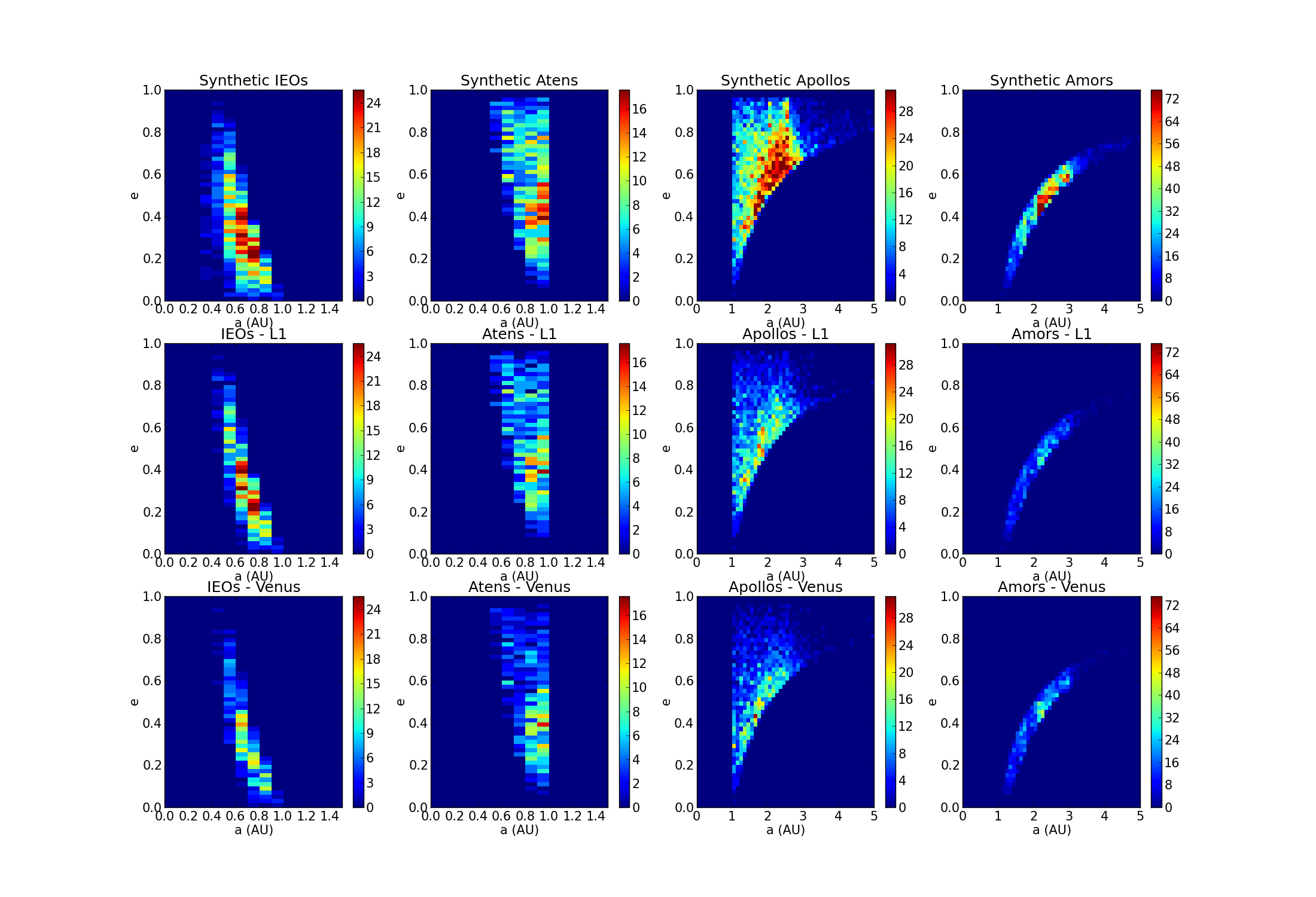}
\caption{\label{fig:a_ecc} The distribution of semi-major axis vs. eccentricity for one of the 25 input synthetic NEA subpopulations (top row), compared with the distributions of objects detected with $>$22 day observational arcs that were found by the L1 survey (middle row) and the Venus-trailing survey (bottom row) after five years. }
\end{figure}

\begin{figure}[H]
\figurenum{16}
\includegraphics[width=6in]{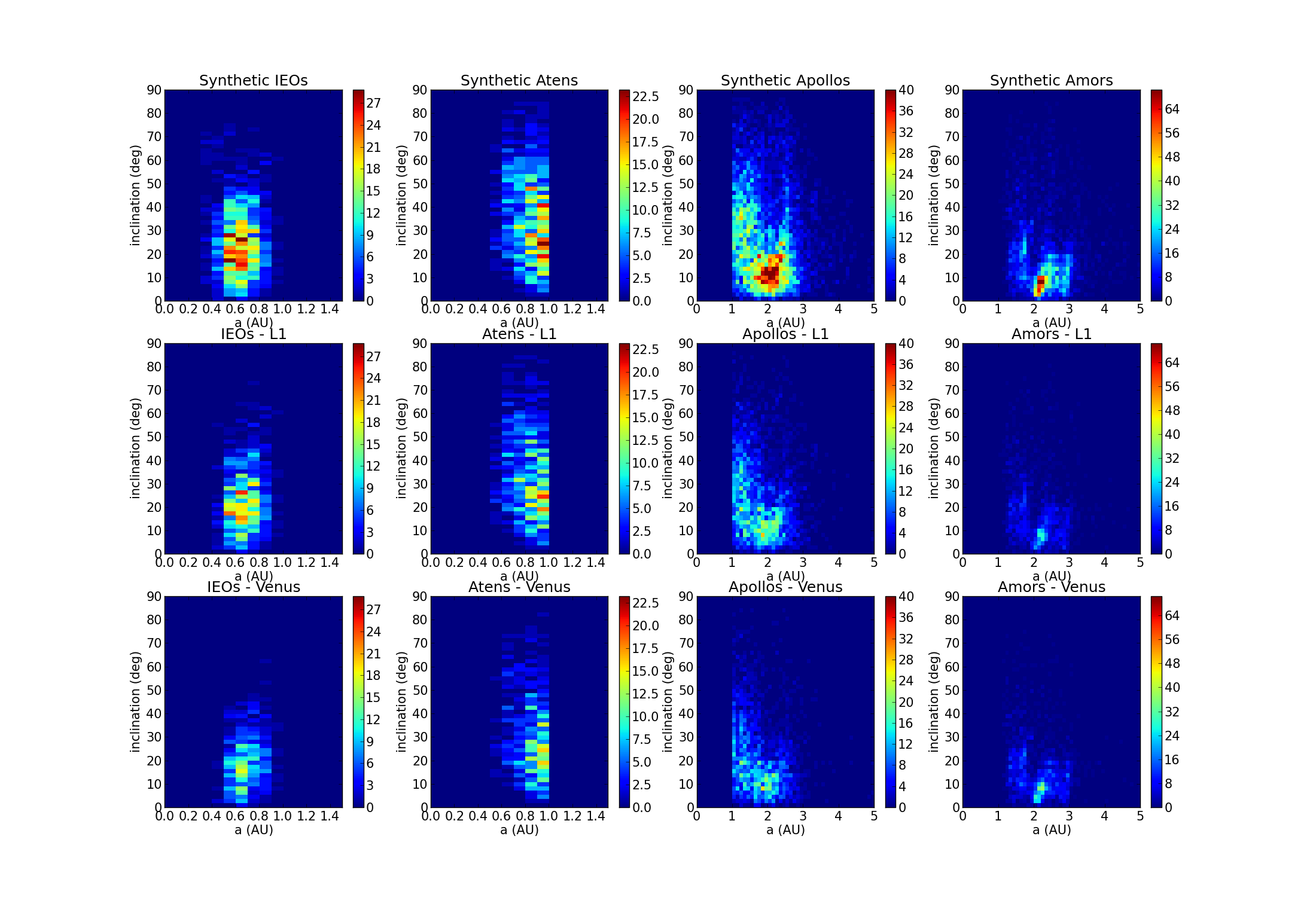}
\caption{\label{fig:a_incl} The distribution of semi-major axis vs. inclination for one of the 25 input synthetic NEA subpopulations (top row), compared with the distributions of objects detected with $>$22 day observational arcs that were found by the L1 survey (middle row) and the Venus-trailing survey (bottom row) after five years.}
\end{figure}

Figures \ref{fig:a_ecc} and \ref{fig:a_incl} show the orbital element distributions for one of the 25 input synthetic populations of IEOs, Atens, Apollos, and Amors, along with the orbital element distributions of the objects detected with $>$22 day observational arcs after five years for the L1 and Venus trailing surveys.    As expected based on the results of Figures \ref{fig:NEATM} and \ref{fig:FRM}, the L1 survey is more efficient at detecting IEOs, Atens, and Apollos, and the Venus-trailing survey is better at detecting Amors.  

\begin{figure}[H]
\figurenum{17}
\includegraphics[width=6in]{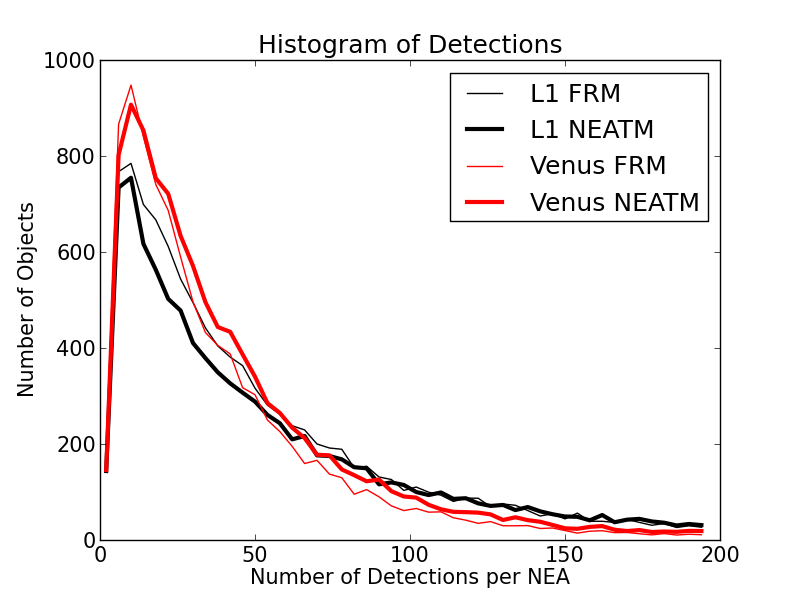}
\caption{\label{fig:ndets} The number of detections per NEA for L1 (black lines; heavy and lighter weight lines represent NEATM and FRM results) and Venus-trailing surveys (red lines) after five years. }
\end{figure}

\begin{figure}[H]
\figurenum{18}
\includegraphics[width=6in]{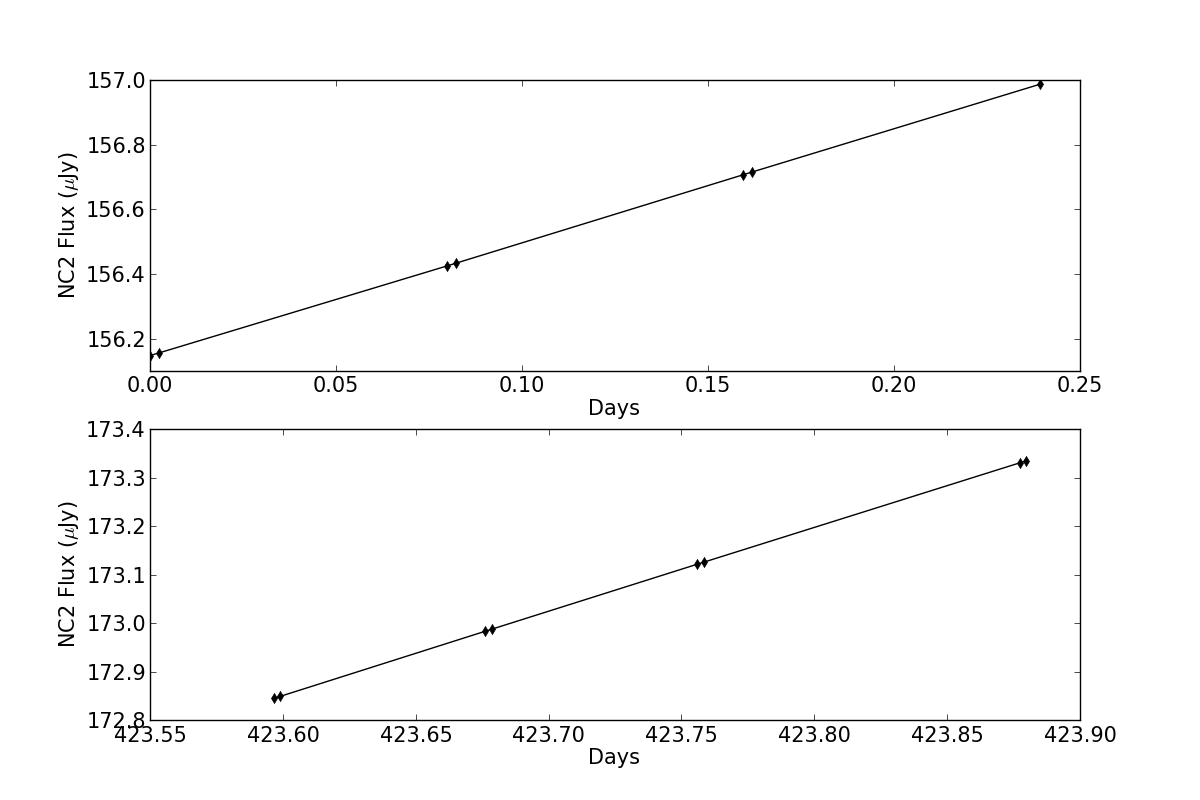}
\caption{\label{fig:lightcurve_S0000oFk} Sample observations of a synthetic Aten detected at two epochs spanning $\sim$24 days (black diamonds); this cases represents a typical object detected by the L1 survey.  }
\end{figure}

\begin{figure}[H]
\figurenum{19}
\includegraphics[width=6in]{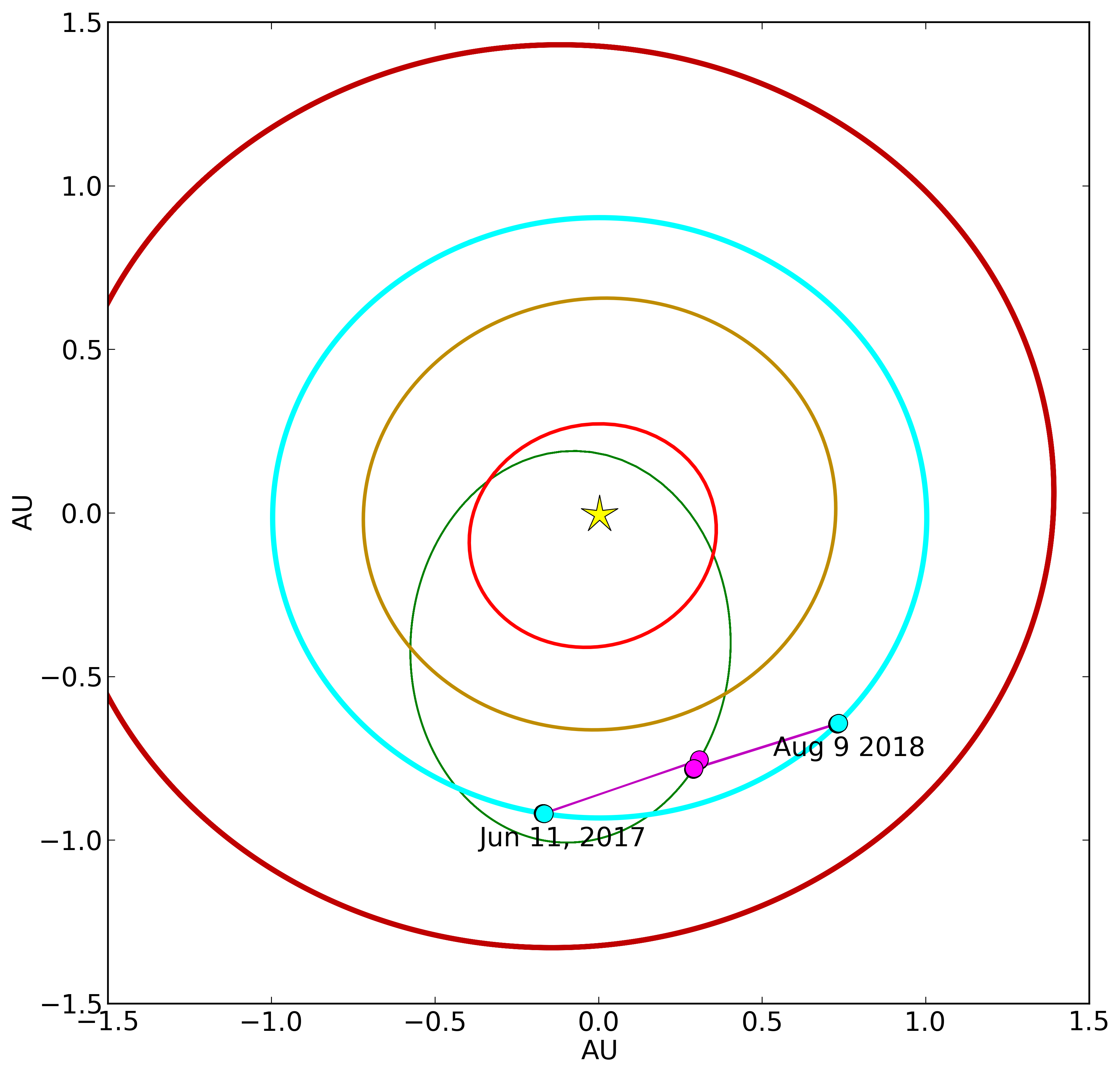}
\caption{\label{fig:geometry_S0000oFk}  Viewing geometry of the epochs at which the Aten shown in Figure \ref{fig:lightcurve_S0000oFk} was detected.  The orbits of Mercury, Venus, and Earth are shown for reference; the positions of Earth and the asteroid are shown as cyan and magenta points, respectively.  The orbit of the Aten is shown in green; magenta lines represent the line-of-sight between the Earth and the asteroid at each epoch.  This object has a 0.58 year orbital period, so it is observed by the L1 survey a nearly the same point in its orbit $\sim$423 days after its first apparition as indicated by the dates shown in the figure, albeit from a different angle.}
\end{figure}

\begin{figure}[H]
\figurenum{20}
\includegraphics[width=6in]{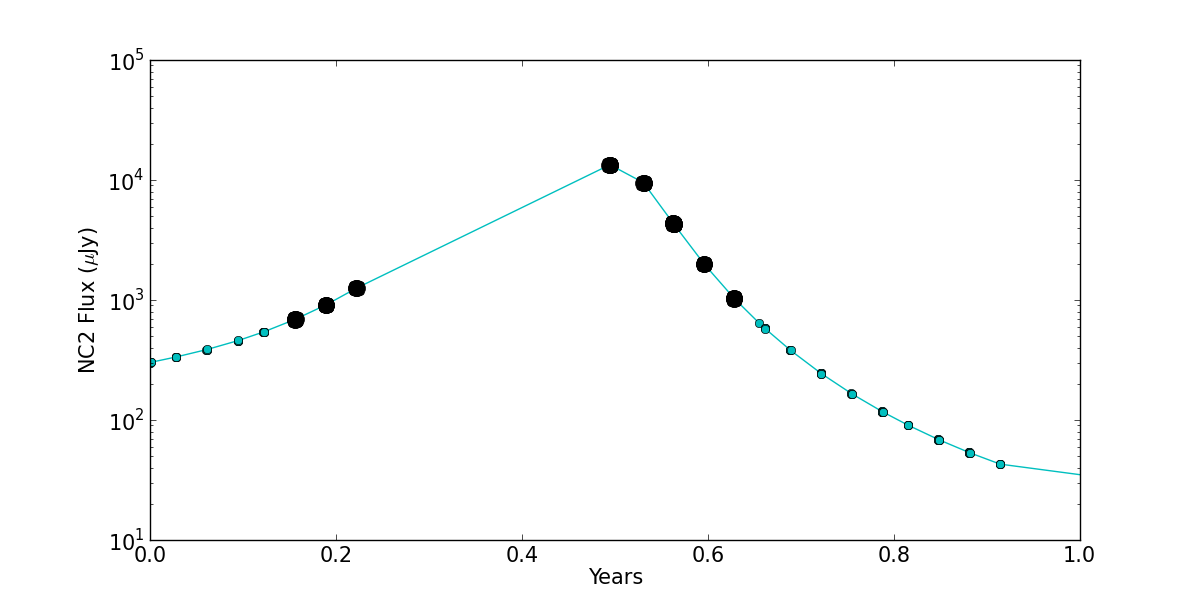}
\caption{\label{fig:lightcurve_S000112i} Sample observations of a synthetic Apollo detected 44 times over $\sim$6 months (black circles; each circle contains $\sim$3-13 individual observations).  The cyan points show the flux of the object when it passes through the L1 survey's field of view but is not bright enough to be detected.  }
\end{figure}

\begin{figure}[H]
\figurenum{21}
\includegraphics[width=6in]{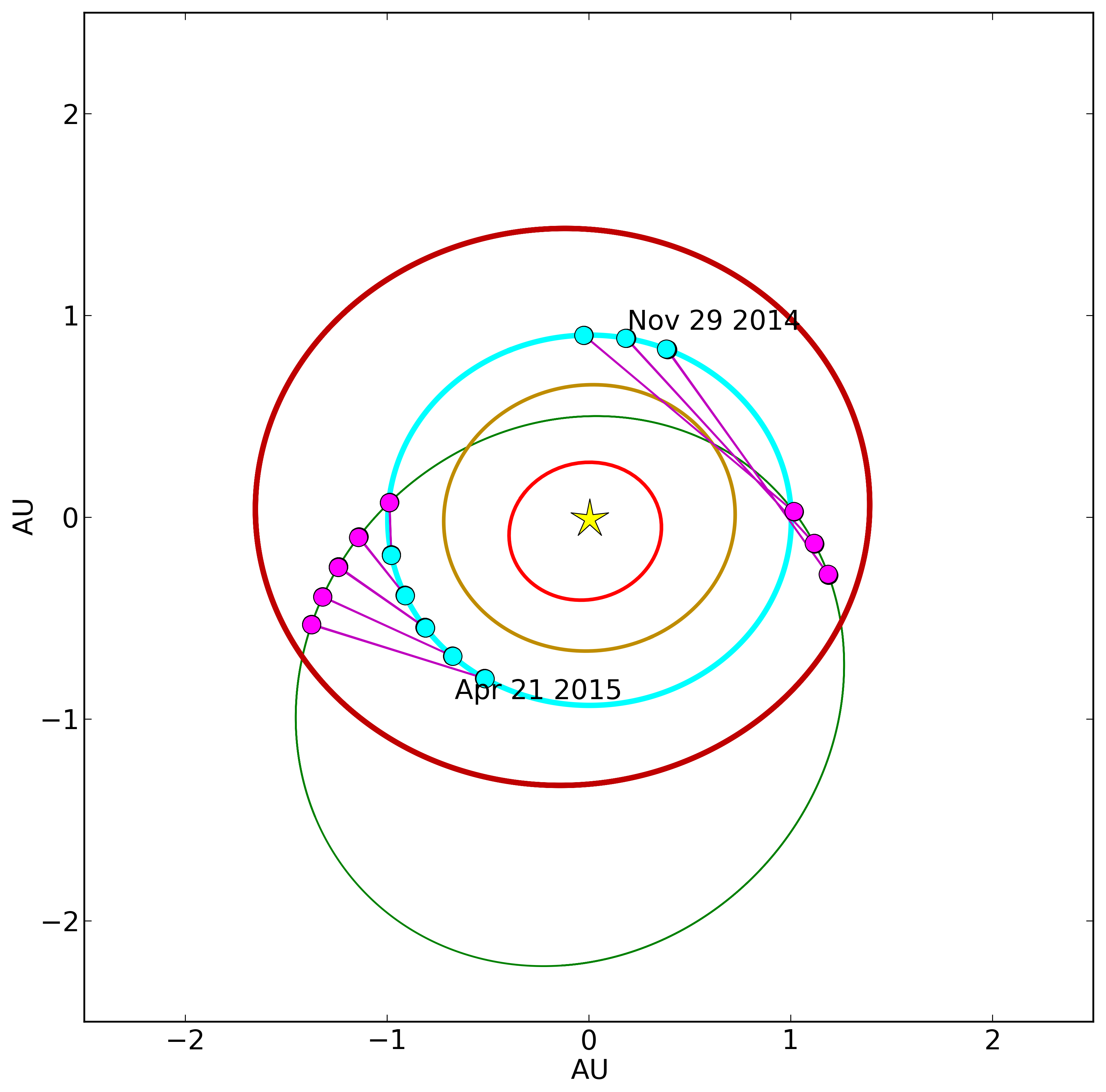}
\caption{\label{fig:geometry_S000112i}  Viewing geometry of the two epochs at which the Apollo shown in Figure \ref{fig:lightcurve_S000112i} was detected.  This object has a 2.3 year orbital period.  Color-coding is the same as Figure \ref{fig:geometry_S0000oFk}. }
\end{figure}

Figure \ref{fig:ndets} shows the average number of detections per object over the course of a six-year survey for both L1 and Venus-trailing surveys.  Many NEAs receive approximately two dozen or more detections, with a large fraction receiving $>$50 collected from multiple viewing geometries.  Figures \ref{fig:lightcurve_S0000oFk} - \ref{fig:geometry_S000112i} show sample lightcurves and viewing geometries for two objects, one with an average number of observations, and one with $>$50 observations.  Most objects are seen at a minimum of two different viewing geometries, which is useful for thermophysical modeling.  

\begin{figure}[H]
\figurenum{22}
\includegraphics[width=6in]{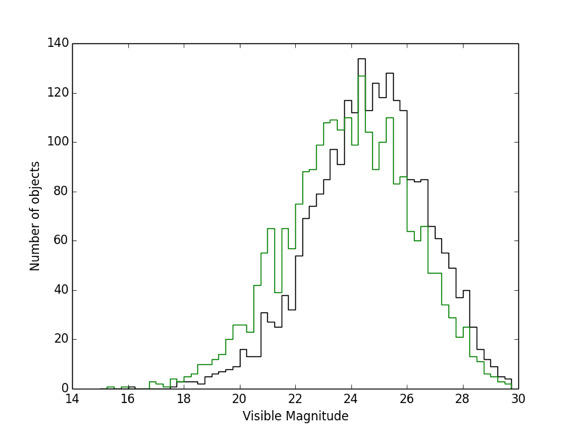}
\caption{\label{fig:visible} The distribution of visible magnitudes for the L1 survey.  The brightness at the time of the first observation for each object is shown (green line), along with the average brightness for each object's tracklets during the first year of the survey (black line).}
\end{figure}

The distribution of visible magnitudes for the L1 survey is shown in Figure \ref{fig:visible}, both at the epoch of the first observation of each object and averaged over each tracklet throughout the first year of the survey.  Visible magnitudes are computed from absolute magnitude $H$ using the geometric visible albedo and diameter assigned to each synthetic object and the relationship \begin{equation}H = -5\cdot log_{10}\left( \frac{D \sqrt{p_{v}}}{1329}\right).  \end{equation} where $D$ is the diameter in kilometers \citep{Bowell.1989a}.  The average visible magnitude of objects detected by the L1 survey is V$\sim$25 mag.

\begin{figure}[H]
\figurenum{23}
\includegraphics[width=6in]{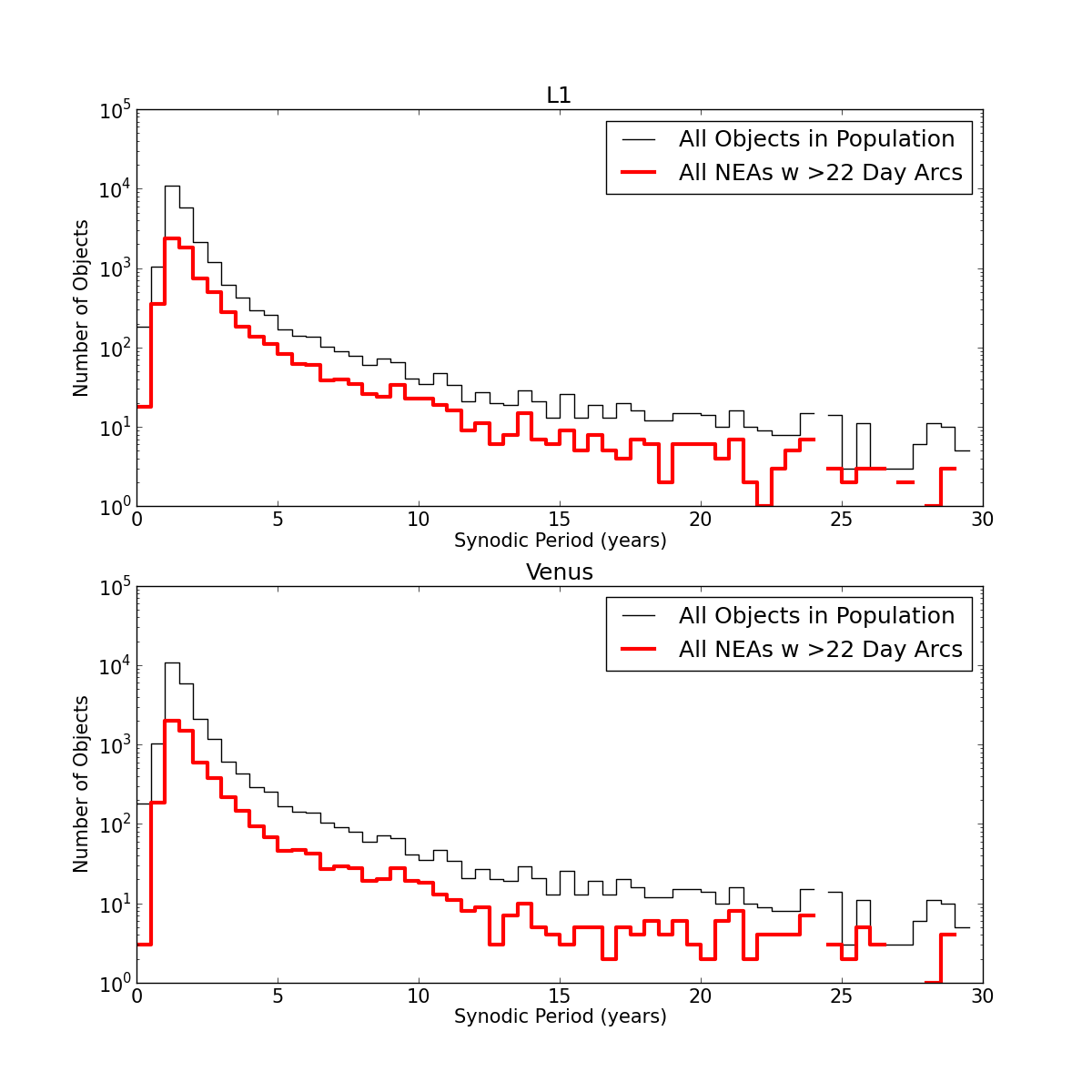}
\caption{\label{fig:synodic_period} The distribution of the absolute values of synodic periods for all NEAs in one of the 25 synthetic populations is shown (black lines) for both L1 (top) and Venus-trailing surveys (bottom).  The red lines indicate the synodic period distribution for all objects found with $>$22 day observational arcs by each survey after five years.}
\end{figure}

Figure \ref{fig:synodic_period} shows the absolute values of the synodic periods of NEAs detected by both L1 and Venus-trailing surveys, where synodic period $S$ measured in years with respect to Earth is defined as \begin{equation}\frac{1}{S} = 1 - \frac{1}{P}. \end{equation}  Figure \ref{fig:synodic_period} shows that the Venus-trailing survey achieves essentially identical performance in terms of the ability to detect NEAs of various synodic periods.  At the integral completeness levels achieved, the survey simulations find no ``hidden" population of NEAs with decades-long synodic periods that are unobservable to the L1 survey, since the NEAs librate into the fields of regard of the L1 survey and do not stay directly behind the Sun.  

The net result is that, even assuming there are no penalties to sensitivity or reliability from the lossy data compression needed from Venus-trailing orbit, the L1 survey outperforms the Venus-trailing survey slightly in terms of integral survey completeness and number of detections per object for NEAs larger than 140 m.

We also studied the case of varying the range of solar elongations for the L1 survey, including scanning through opposition (which requires moving the spacecraft to L2 to avoid viewing the Earth).  Assuming that such a survey covers sky at the same rate and must carry out the same cadence, surveying through opposition from 110$^{\circ}$ solar elongation detects only 34\% of PHAs after six years and no IEOs.  This poor result is not surprising, given that \citet{Chesley.2004a} found that PHAs and potential impactors are preferentially found at lower solar elongations.  Moreover, surveying through opposition presents serious engineering obstacles in terms of keeping sunlight off sensitive thermal and optical surfaces.  

There are several areas for improvement in the simulation.  First, the synthetic source lists created for use with XMOPS did not include models of spurious detections.  Experience with WMOPS has shown that unidentified artifacts can severely impact tracklet reliability, including those artifacts associated with bright moving objects (e.g. Jupiter and Mars).  Artifact models based on the HgCdTe arrays used by WISE could be created and injected along with models of spurious detections such as diffraction spikes and ghosts.  Second, the XMOPS portion of the simulation could be extended from its current $\sim$260 square degree footprint spanning two years to cover the entire sky over a longer duration.  Third, the population of co-orbitals such as Earth Trojans and NEAs in so-called ``horseshoe" orbits must be estimated and included in population models.  Finally, it may be possible to develop a survey cadence that is better optimized for NEA detection and characterization.  The cadence described here represents an initial result that could potentially be further optimized.  

\subsection{Orbit Determination}
A critical parameter in determining the success of the trial cadence tested for both surveys is the ability to compute reliable orbits from the simulated observations.  To that end, we evaluated the orbits fit to a set of candidate tracklets resulting from both L1 and Venus-trailing surveys.  The determination of the success or failure of any asteroid observational cadence is relatively straightforward.  Objects must be linked with a high enough reliability and observed over an arc of sufficient length that the vast majority of the objects can be identified at subsequent appearances.  Orbit fits to a set of candidate tracklets resulting from both L1 and Venus-trailing surveys using the standard, existing tools employed for real-time linking by the MPC were evaluated.  

The method is as follows: First, a \Vaisala\ orbit is computed for each tracklet.  The \Vaisala\ method  is useful for estimating an orbit from a short-arc set of observations spanning a day or two \citep{Vaisala.1939a}.  The technique assumes that the object is at apsis at either the starting or ending observation, so that the object's instantaneous radial velocity is zero. It can provide a useful discriminator between Main Belt and NEA orbits when very few observations are in hand. Next, an attempt is made to find additional tracklets for that object by comparing residual positional differences between each orbit and tracklet pair (noting magnitude and consistency of astrometric residuals).  Then, orbits are computed using Gauss' method \citep[or other methods such as those described in, e.g.,][]{Marsden.1985a} for linked tracklets.  The process is repeated using a different starting orbit technique \citep[c.f. the assumed elements or generalized \Vaisala\ techniques discussed in ][]{Marsden.1991a}.  At each step, confirmed linkages result in tracklets being removed from the tracklet stack, further simplifying linkage attempts going forward.  

Approximately 10,000 tracklets from the synthetic all-sky survey were evaluated in this fashion.  In the procedure described above, all objects were treated as new and unidentified.  In practice, this is an unrealistic assumption since the current MBA catalog contains $\sim$500,000 objects with well-determined orbits.  By the time a new space-based survey mission could be launched, even assuming construction started today, we expect the MBA catalog to contain $\sim$1 million objects with well-determined orbits, and the NEO catalog could be expected to contain perhaps 15,000 - 20,000 objects.  As a standard practice, the MPC runs all tracklets through procedures to check against known objects to remove in real time the easiest linkages to make.  It is expected that $>$30\% of tracklets will be immediately identifiable, greatly simplifying the linking process. This result also serves as a warning that any dramatic increase in false tracklet generation rates will greatly complicate linking and thus reliability. It is imperative that the submitted tracklets be dominated by real objects.

\section{Conclusions}
We have constructed a set of simulations that account for updated NEA population models, including physical properties, orbital elements, and numbers.  A portion of our model demonstrates the moving object pipeline's ability to extract and link tracklets in the presence of potentially confusing astrophysical sources as well as other asteroids that are not NEAs.  We have successfully computed accurate orbits for the assumed trial survey cadence using the standard tools employed by the MPC.  However, we note that the ability to compute orbits depends critically on excluding most false detections from the submitted tracklets.  Construction of a more complete catalog of potentially confusing non-NEAs such as MBAs and Jovian Trojans is essential.

We compared the performance of the Venus-trailing and L1 surveys and found that the Venus-trailing survey detects slightly fewer PHAs $>$140 m than the L1 survey, even assuming that no degradation results from the inability to downlink full-frame images.  While the Venus-trailing survey discovers more Amors than the L1 survey, these objects are less likely to constitute impact hazards compared to Atens and Apollos, nor are they likely to be suitable targets for future missions.  The L1 survey discovers more IEOs, Atens, and Apollos than the Venus-trailing survey.  These results demonstrate that the cost, complexity, and risk associated with sending a survey telescope to a Venus-trailing orbit is unwarranted.  While neither survey is capable of fulfilling the 2005 Congressional mandate to NASA to find 90\% of all near-Earth objects larger than 140 m in diameter by 2020, an advanced space-based survey can make significant progress quickly.  

No asteroid survey has demonstrated the ability to detect NEOs at low SNR and high reliability using only the equivalent of a spacecraft processor and memory to perform all source extraction algorithms.  The loss of ancillary science and the ability to perform optimal image reprocessing with calibration products derived later in the mission associated with downlinking only regions of interest instead of full-frame images represents an additional penalty for interior-orbiting missions.

These simulations serve as a starting point for optimizing the performance of an advanced survey aiming to largely characterize the hazard posed by NEOs, identify the best targets for future exploration, and deliver a high-quality dataset ripe for solar system science to the community.  Further improvements, such as those described above, can enhance the fidelity of the results.   

\section{Acknowledgments}

\acknowledgments{This publication makes use of data products from the Wide-field Infrared Survey Explorer, which is a joint project of the University of California, Los Angeles, and the Jet Propulsion Laboratory/California Institute of Technology, and NEOWISE, which is a project of the Jet Propulsion Laboratory/California Institute of Technology. WISE and NEOWISE are funded by the National Aeronautics and Space Administration.   We thank the referee, Dr. Alan Harris of Pasadena, for helpful suggestions that greatly improved the manuscript.  This research has made use of the NASA/IPAC Infrared Science Archive, which is operated by the Jet Propulsion Laboratory, California Institute of Technology, under contract with the National Aeronautics and Space Administration. This work is based [in part] on observations made with the Spitzer Space Telescope, which is operated by the Jet Propulsion Laboratory, California Institute of Technology under a contract with NASA.   }

  \clearpage

 \clearpage

\end{document}